\documentclass[acmlarge]{acmart}
\usepackage{CJKutf8}
\usepackage{amsmath}  
\usepackage{tipa}

\usepackage{url} 
\usepackage{hyperref}

\AtBeginDocument{%
  }

\begin{document}

\title{Plant-Centric Metaverse: A Biocentric-Creation Framework for Ecological Art and Digital Symbiosis}

\author{Ze Gao}
\affiliation{%
\institution{Cyanpuppets \& Hong Kong Polytechnic University}
\city{Guangzhou \& Hong Kong SAR} 
\country{China}}
\email{zegao@cyanpuppets.com}

\author{Mengyao Guo}
\affiliation{%
\institution{Harbin Institute of Technology, Shenzhen}
\city{Shenzhen} 
\country{China}}
\authornote{Mengyao Guo is the corresponding author.}
\email{guomengyao@hit.edu.cn}

\author{Zheng Wang}
\affiliation{%
\institution{Nanyang Technological University}
\city{Singapore} 
\country{Singapore}}
\email{wang1796@e.ntu.edu.sg}

\author{Xiaolin Zhang}
\affiliation{%
\institution{University of Auckland}
\city{Auckland} 
\country{New Zealand}}
\email{xiaolinzhang1001@gmail.com}

\author{Sihuang Man}
\affiliation{%
\institution{Columbia University}
\city{New York} 
\country{United States}}
\email{sm8497@nyu.edu}

\renewcommand{\shortauthors}{Gao and Guo et al.}

\begin{abstract}
\textbf{Digital ecological art} represents an emergent frontier where biological media converge with virtual environments. This study examines the paradigm shift from \textbf{anthropocentric} to \textbf{plant-centered artistic narratives} within the \textbf{metaverse}, contextualizing how digital platforms transform ecological expression. However, current frameworks fail to systematically guide artists in leveraging \textbf{plant agency} for \textbf{digital symbiosis} that transcends human-centered creation. We propose the \textbf{Biocentric-Creation Transformation Ideology (BCTI)} framework and validate it through \textbf{multimodal case studies} spanning bio-art, NFTs, and VR ecosystems (2013-2023). Our analysis reveals: (1) \textbf{Metaverse ecosystems} enable unprecedented \textbf{plant-algorithm co-creation}, with biological artworks increasing by 133\% in premier archives (2020 vs 2013); (2) \textbf{Digital symbiosis} manifests through blockchain DAOs where plants govern human-plant collaborations; (3) \textbf{Algorithmic photosynthesis} in VR environments reshapes ecological aesthetics through real-time biodata translation. The BCTI framework advances \textbf{ecological art theory} by systematizing the transition from representation to \textbf{plant-centered agency}, offering artists a blueprint for \textbf{post-anthropocene creation}. This redefines environmental consciousness in virtual realms while establishing new protocols for \textbf{cross-species digital collaboration}.
\end{abstract}


\maketitle

\section{Introduction}

Biological art, which often integrates living organisms as part of the artwork, is a potent medium through which the principles of deanthropocentrism can be explored and expressed~\cite{weintraub2012life}. Relating deanthropocentrism to our discussion on the evolution of biological art and the Biocentric-Creation Transformation Ideology (BCTI), this approach plays a critical role in shaping the ideologies and practices within this artistic field. This paper thoroughly examines the trajectory within biological art, mapping its transition through the lens of the BCTI. In this paper, we introduced a systematic framework that embeds ecological tenets and biocentric viewpoints within the sphere of artistic creation. This framework emphasizes that artistic creation should transcend human subjectivity, incorporating the integrity of the ecosystem, the intrinsic value and agency of non-human life (especially plants), and the responsibility of humans as a part of the ecosystem into its core considerations. By guiding artists to focus on life processes (such as plant growth)~\cite{guo2025slimo}, ecological interrelations (such as symbiosis), and environmental ethics, this framework provides a theoretical foundation and practical path for artists to shift from a human-centered narrative to an ecosystem-centered perspective.

The transformation ideology outlined in this paper acts as a comprehensive guide for biological artists, charting a course for incorporating biological and ecological components into their art. It provides artists with a multidisciplinary understanding that spans social, biological, semiotic, and aesthetic realms, empowering them to imbue their works with rich, integrative layers of meaning. This ideological and practical scaffold not only fosters artistic innovation but also encourages the reflection of complex ecological interrelations within the artistic process.

\subsection{Rise of Ecological Art and Plant-centric Approaches}

The rise of ecological art and plant-centric approaches marks a pivotal shift in the art world, reflecting a growing awareness of environmental issues and the interconnectedness of all life forms. This movement sees artists exploring the complex relationships between humans and their natural environment. The rise of ecological art is an early practical exploration of the BCTI. Through land art, environmental intervention, and other means, artists question the binary division between art and life, as well as between humans and nature, and view art practice itself as a process of reflecting on the impact of human activities on the environment and attempting to establish a dialogue with the ecosystem.

\begin{figure}
\includegraphics[width=\textwidth]{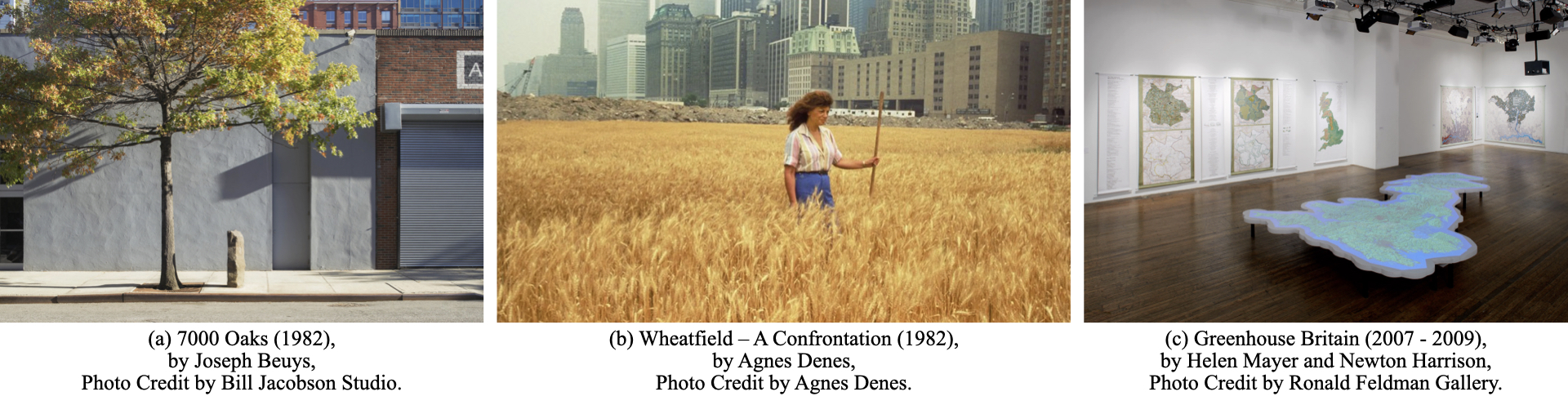}
\caption{Early Works of Ecological Art.
\copyright Artist Mentioned.}
\Description{Early Works of Ecological Art.}
\label{intro2}
\end{figure}

Ecological art, or ``Eco Art,"  as a profound response to the increasingly severe environmental crisis and the consequences of modernity, emerged in the 1960s and 1970s amid the growing environmental protection movement. Its cultural impetus is rooted in the convergence of multiple ideological trends: the rise of ecology as a discipline and the infiltration of its systems perspective into public consciousness; critical reflection on the overexploitation of nature by industrial civilization and the resulting ``Silent Spring"; the counterculture's questioning of the values and lifestyle of industrial society; and the criticism of art's detachment from social reality and confinement within the gallery system. Against this backdrop, artists began to question the constructed barriers between art and life, humans and nature, and actively expanded their artistic practices into a broader ecological context, directly engaging with environmental issues and reflecting on the impact of human activities on the Earth. For example, the 1960s and 1970s marked the early symbolic and interventionist periods of ecological art. Joseph Beuys'``7000 Oaks" (see Figure~\ref{intro2}a) and Agnes Denes's ``Wheatfield - A Confrontation" (see Figure~\ref{intro2}b) represent the early use of plants in ecological art. Beuys regarded the planting of oak trees as a form of social sculpture and a promise for the future; the trees became symbolic living entities connecting society and the environment, and their growth process itself was part of the art. Denes, on the other hand, sown a wheat field in the city center, using plant life as a direct and temporary critical tool to expose issues with urban land policies. Here, plants served as temporal witnesses and protest media.

Intertwined closely with ecological art and often serving as one of its key practice paths is the plant-centered artistic approach. This kind of practice particularly focuses on plant life itself, referring not to it simply as a part of the environment but as a core subject and cooperative medium with unique ways of perception, communication, and existence. The plants are not only depicted, but also integral components of artworks. Such practices are diverse, from the integration of live plants in installations and performances to the use of botanical materials in sculptures and textiles. Artists may cultivate living sculptures in gardens, collaborate with plants to create art, or employ scientific techniques to create living, breathing biological art. 

\begin{figure}
\includegraphics[width=\textwidth]{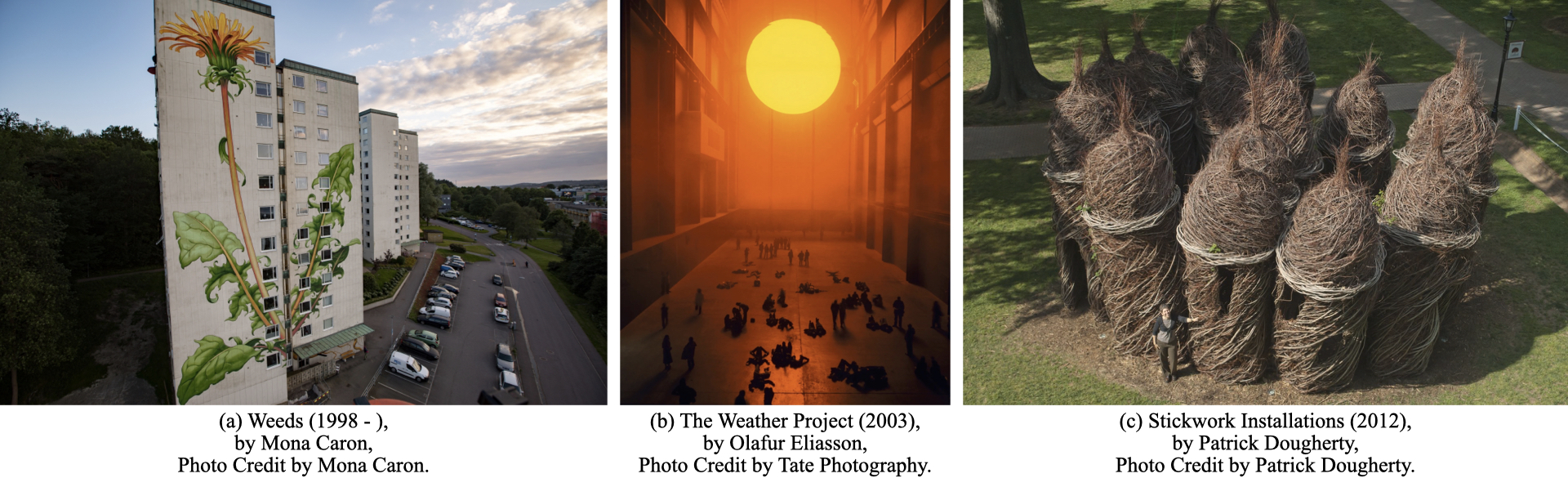}
\caption{Artworks of Ecological Art.
\copyright Artists Mentioned.}
\Description{Artworks of Ecological Art.}
\label{intro22}
\end{figure}

In the last decade of the 20th century, artistic creation with plants as the theme and medium entered a period of ecological process simulation and perception exploration. Helen and Newton Harrison's ``Greenhouse Britain" (see Figure~\ref{intro2}c) simulated the impact of climate change on plant distribution through complex maps and information visualization, taking future changes in plant ecological niches as the core narrative content, which reflected a systematic concern for the interdependent relationship between plants and the environment. Olafur Eliasson's ``The Weather Project" (see Figure~\ref{intro22}b), although named after a meteorological phenomenon, used the humid misty environment under its huge artificial sun and the experience it triggered to ingeniously guide the audience to reflect on their perception and responsibility towards the climate (and the plant life that supports it). Although the plants were not directly present, the environmental conditions on which their survival depends became the focus of perception. Patrick Dougherty's ``Stickwork Installations" (see Figure~\ref{intro22} c) directly used felled plant materials (branches) to create large-scale temporary landscape structures through weaving techniques, emphasizing the organic forms of natural materials, the life cycle (from growth to decay), and the wisdom of human-nature collaboration in craftsmanship.

Since the twentieth century, urban ecological participation has been more closely integrated with the agency of plants. Mona Caron's ongoing ``Weeds" project (see Figure~\ref{intro22}a) features the tenacious weeds growing in the cracks of cities, magnifying, beautifying, and heroizing them through large-scale murals. This is not merely a depiction of plants; it is an artistic endowment of visibility and subjectivity to the overlooked urban flora, inspiring viewers to re-examine urban ecological resilience and the value of non-human life, directly embodying a ``plant-centered" narrative perspective.

Plant-centric art frequently blurs the boundary between art and science and that between landscape and architecture. It includes the study of plant behavior and intelligence, investigating plant communication, and using plants to comment on issues such as genetic modification, biodiversity loss, and global warming~\cite{calvo2020plants}. This plant-centered artistic approach can be regarded as a deeperened manifestation of the ``biocentric-creation transformation ideology (BCTI)" framework in a specific medium. It not only depicts plants, but also regards them as collaborators and narrators with their own life logic and temporality, challenging the traditional notion of plants as purely passive resources and promoting a further shift in artistic creation from anthropocentrism to ecocentrism.

As the world grapples with the realities of the Anthropocene and its attendant ecological crises, ecological art and plant-centric approaches act as powerful change agents, inspiring a collective rethinking of humanity's place in the natural order and our impact on the planet we live on.

\subsection{The Evolution of Art and Technology}

The evolution of artistic media and tools highlights the profound interaction between human creativity and technological development. From the use of natural pigments in prehistoric cave paintings to the popularization of oil paints during the Renaissance, and then to the advent of photography and acrylic paints brought about by the Industrial Revolution, each technological breakthrough has expanded the boundaries of artistic expression. The arrival of the digital revolution has completely reshaped the creative paradigm: digital painting software offers unprecedented texture simulation capabilities, 3D printing technology allows the materialization of virtual models, and computer-generated images have opened up a new visual dimension.~\cite{hertzmann2018can}~\cite{moss2019expanded} These technologies not only transform the form of art but also redefine the boundary between art and reality.

The liberation and democratization of space have broken the barriers of both the physical and cognitive realms. The rise of the Internet has caused a fundamental transformation in the art ecosystem. Physical restrictions on the creation and dissemination of art have been completely shattered, making it possible for global exchanges of ideas, collaborative creation, and instant sharing~\cite{bolter2021reality}. Digital art, net art, and virtual installations have allowed artists to bypass traditional institutions and directly connect with a global audience. As the democratization of art accelerates, the trend of decentralization becomes increasingly prominent - the subject of creation has expanded from professional artists to a broad community, and the power structure of art has undergone profound reconstruction. This spatial liberation is not only reflected in the physical aspect, but also extends to the cognitive domain: art is no longer confined to visual forms, but integrates auditory, tactile, and other multisensory channels to create immersive cross-media experiences.

In the diachronic evolution of artistic forms, technological innovations have constantly given rise to new art genres. In 1964, Claude Beylie defined comics as the ninth art form~\cite{beylie1964bande}; nowadays, with the maturation of virtual reality technology and the construction of the metaverse ecosystem, the tenth art form has emerged. Virtual communities such as Decentraland, Sandbox, and Cryptovoxels have become experimental fields for new forms of artistic expression (see Figure~\ref{intro2}), where artificial intelligence art, generative art, and bioart interweave and coexist. The leap in communication and imaging technologies has enabled art to completely break free from the constraints of physical carriers and build a self-consistent aesthetic system in the digital world. It can be said that the tenth form of art is a significant marker of the establishment of the metaverse as a new era of art.

When art shifts from physical spaces to virtual domains, its social functions also need to be reexamined. Against the backdrop of an increasingly severe ecological crisis, the value orientation of art undergoes an essential transformation - from pursuing formal autonomy to taking on environmental responsibilities. The flourishing of Eco Art is precisely a response to this: its etymological root ``oikos" (meaning ``home" in ancient Greek) indicates a profound connection between artistic creation and the ecosystem. This functional repositioning prompts artists to rethink the essence of creation: art is not only a projection of human spirit but should also serve as a medium to mend the rift between humans and nature.

\begin{figure}
\includegraphics[width=\textwidth]{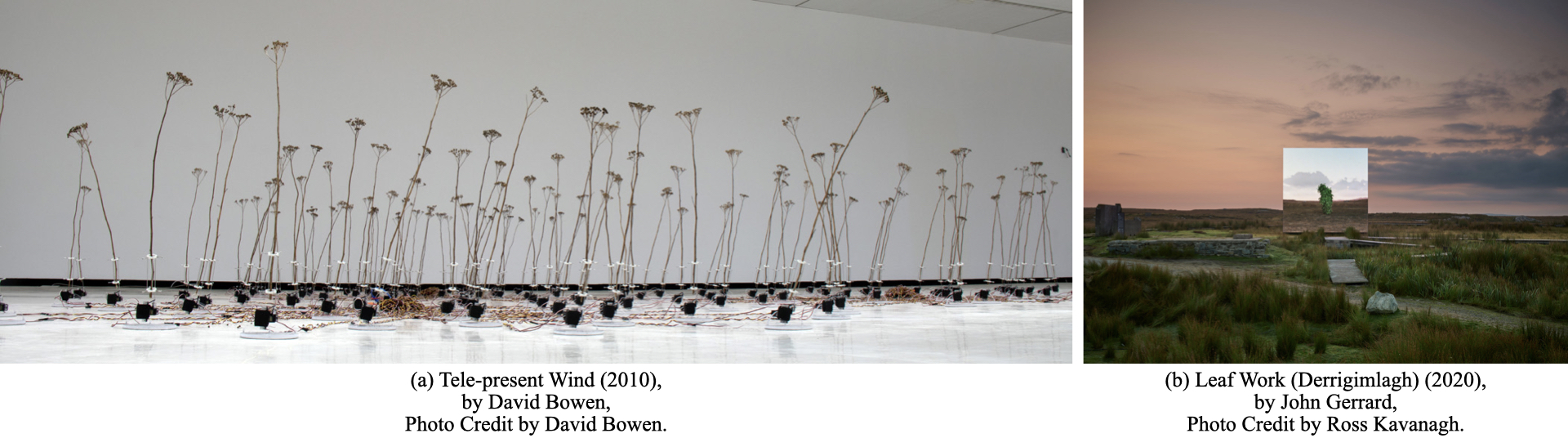}
\caption{Artworks of Ecological Art.
\copyright Artists Mentioned.}
\Description{Artworks of Ecological Art.}
\label{intro4}
\end{figure}

The metaverse offers an unprecedented space for ecological art expression. Artists simulate natural processes through digital media or generate dynamic works using environmental data, reconstructing ecological consciousness in virtual fields. David Bowen's ``Tele-present Wind" (see Figure~\ref{intro4}a) makes the intangible wind tangible by driving the device's movement with real-time wind speed data; John Gerrard's analog landscape ``LeafWork" (see Figure~\ref{intro4}b) precisely simulates the photosynthesis of leaves with algorithms, revealing the exquisite rhythms of plant life. These creations are not only explorations of technological potential, but also poetic interpretations of ecological connections.

The ecological turn of the artistic function is bound to trigger a reconfiguration of the identity of the creative subject. In the framework of the tenth art, the artist transforms from an individual creator to a coordinator of a cross-disciplinary collaboration network: (1) The shift from creator to collaborator. The ecological art in the metaverse requires cross-disciplinary cooperation. (2) The transformation from a human spokesperson to an ecological translator. The artist builds an interface for dialogue between plants and technology. This transformation is a profound manifestation of deanthropocentrism - the artist cedes part of the control, allowing plants to become co-creators of art. This shift holds a dual significance in the dimension of deanthropocentrism: it not only breaks the monopoly of the artist's authority but also facilitates creative dialogue between humans and non-humans.

In general, the integration of art and technology ultimately leads to a deep practice of dehumanization. Technology becomes a channel for non-human life to express its agency. This creative paradigm dissolves the central position of the human subject, establishing a cyclical relationship where human creation nurtures plants, plant growth enriches human experience, and plant communities jointly shape social structures, thus constructing a new aesthetic ethics of life co-construction in the era of the metaverse.

\subsection{Ecological Aesthetics and Art Forms: From Theoretical Criticism to Life Practice}

Frederick Steiner's theory of ecological aesthetics has overturned the traditional art paradigm. He criticized anthropocentric aesthetics and questioned the ``Fine Arts" tradition, which confines ``beauty" to human sensory experiences. Moreover, he redefined the essence of ecological art and proposed the concept of Eco Art, emphasizing that art should embody ecological principles (e.g., material cycling and energy flow), providing a new perspective for art that transcends the traditional ``Fine Arts" and encompasses Eco Art, that is, the art of ecology~\cite{steiner2019toward}. In addition, Steiner integrated land ethics, inherited Aldo Leopold's thoughts, and advocated that ecosystem services (e.g., water conservation and carbon sequestration) have intrinsic aesthetic value.

Ecological aesthetics promotes the reconstruction of the art value system, transforming the traditional static appreciation mode of basic aesthetics into a dynamic participatory process of symbiotic aesthetics. This is specifically reflected in the following three aspects: (1) Subject displacement. Humans have shifted from aesthetic dominators to ecological co-constructors. (2) Value re-evaluation. Life processes such as plant growth and microbial metabolism are endowed with an independent aesthetic status. (3) Ethical shift. Using the resilience of life to counter the narrative of human violence.

With the development of the digital revolution, ecological aesthetics has also permeated the virtual world. Digital art can simulate natural processes or create dynamic works that reflect ecological patterns and systems by using environmental data. It can be said that digital art has expanded the dimension of ecological expression of the virtual field. David Bowen's data-driven installation ``Tele-present Wind" (see Figure~\ref{intro4}a) converts real-time wind speed into mechanical movement, making the invisible climate embodied, which reflects the process of data materialization. John Gerrard's analog landscape ``LeafWork" (see Figure~\ref{intro4}b) simulates the photosynthesis of 100,000 leaves through a real-time rendering engine and builds a digital plant community, which promotes the construction of algorithmic life forms.

As the ultimate practice of ecological aesthetics, biological art can serve as a complete carrier of ecological aesthetic expression. This is mainly due to the significant breakthroughs achieved by biological art in three dimensions: (1) biological art has completely overturned the medium essence of traditional art. The traditional art paradigm relies on inanimate materials, such as pigments and stones, and the material attributes are fixed after the creation is completed. However, biological art takes living organisms (plants, bacteria, fungi) as autonomous creative subjects, and their life processes themselves become the artistic language. (2) Biological art deconstructs the temporal logic of artistic creation. Under the traditional art paradigm, a work is declared complete when the artist finishes painting or sculpting, and the temporal dimension is frozen. Biological art continuously reshapes the form and meaning of the work through the growth, decay, and adaptation of the organism, which presents an open-ended process. (3) Biological art reconstructs the power structure of artistic production. In traditional art, the artist achieves the materialization of personal will through absolute control. Biological art establishes an equal collaboration between humans and non-human life, acknowledging the perception and decision-making power of plants/microorganisms. Plant neurobiology endows creativity with subjectivity, and technology becomes an extension tool of the will of life.

These three breakthroughs jointly promote the realization of the ``biocentric creation and transformation concept": the activation of media subverts the worship of the materiality of art objects; the temporality of growth deconstructs human monopoly over the creative process; and the ethics of symbiosis reconstructs the equal relationship across species.

\begin{figure}
\includegraphics[width=\textwidth]{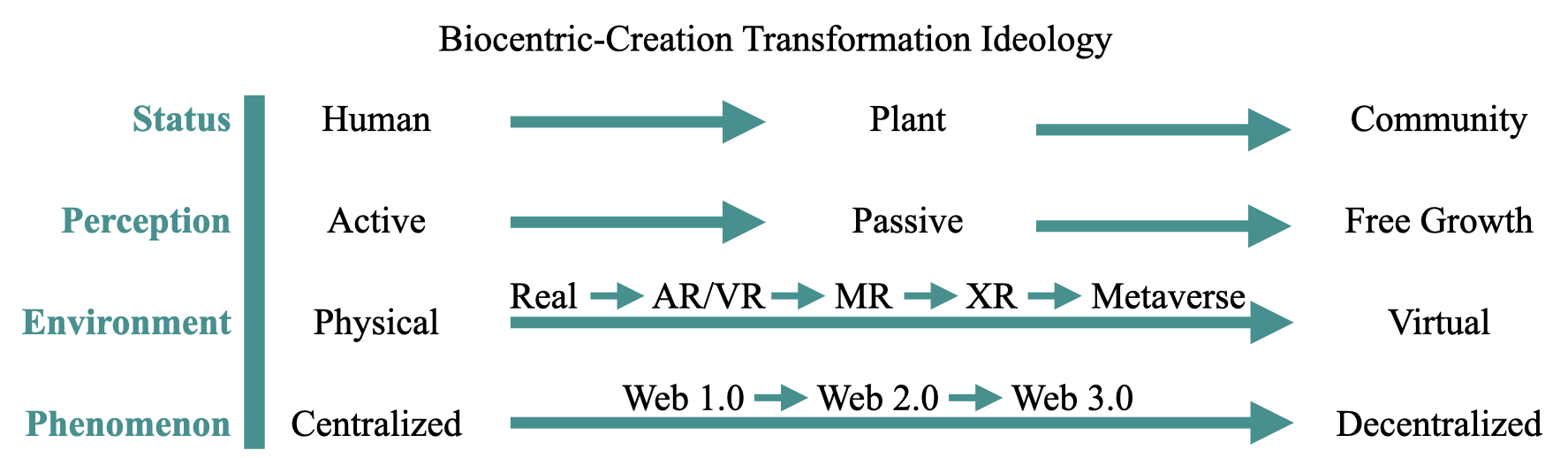}
\caption{The ``Biocentric-Creation Transformation Ideology".
\copyright Credit by Authors.}
\Description{The ``Biocentric-Creation Transformation Ideology".}
\label{gao}
\end{figure}

The BCTI is a systematic framework proposed in this paper (see Figure~\ref{gao}). Its essence lies in establishing an ecological symbiotic cycle of human creation, plant growth, and social response, subverting the unidirectional control model of artistic creation. In the metaverse field, BCTI promotes the upgrading of dehumanization. Digital plant agents represent the life of plants, making plant perception the behavioral guideline for humans. Blockchain technology enables the automation of BCTI, which has led to the formation of the first ``human-plant" co-governed DAO community. BCTI achieves dehumanization through the transfer of power by the creative subject and the ecological reset in the time dimension. Under the mechanism of BCTI, plants become co-creators, and technology transforms from a tool controlled by humans to an extension medium of plant will. Moreover, BCTI defines dynamic temporality through plant growth, forcing humans to accept the dominance of life rhythms over the artistic process. Plants become the hub of transformation, connecting virtual actions with real ecological restoration. Thus, plant-centered ecological art helps trigger collective ecological practices.

Contrary to the traditional view that biological art must be confined to tangible, real-world interactions, our research expands the possibilities into the digital domain. We propose continuing and expanding biological art within digital spaces such as the Metaverse and through AI-related artworks. The research purpose of this study is to explore the paradigm shift of art narrative from anthropocentrism to non-anthropocentrism, with a particular focus on how plant-centered artistic expressions in the metaverse era can achieve this transformation through the framework of BCTI. Specifically, the research aims to analyze the historical evolution of bioart, propose and verify a systematic framework to guide artists to adopt an ecosystem-centered creative perspective, thereby carrying, enhancing, and disseminating new dimensions of ecological participation in the digital environment. Ultimately, this research is committed to promoting the transformation of art practice from human dominance to ecological collaboration through the integration of virtual reality and to cultivating a broader sense of environmental protection and responsibility.

\section{literature Review}

Many theories contribute to biological art, providing a thoughtful foundation for art exploration. The literature review section meticulously examines the confluence of biological art, cultural anthropology, and ecological thought, delineating how these disciplines collectively inform our understanding of art's capacity to engage with and reflect upon the complex tapestry of human behaviors, cultural expressions, and environmental interactions. Within the realm of cultural anthropology, theoretical lenses such as structuralism, biopolitics, and gender performativity are instrumental in dissecting how artistic expression both mirrors and challenges prevailing cultural norms. Moving into the biological sphere, theories of plant growth, biosemiotics, and ecological aesthetics provide a rich backdrop for artworks that not only foreground the centrality of plants but also delve into the nuances of plant communication, the concept of unintelligence, and the artistic portrayal of environmental equilibrium. 

\subsection{Cultural Anthropology Perspectives on Art}

\begin{figure}
\includegraphics[width=\textwidth]{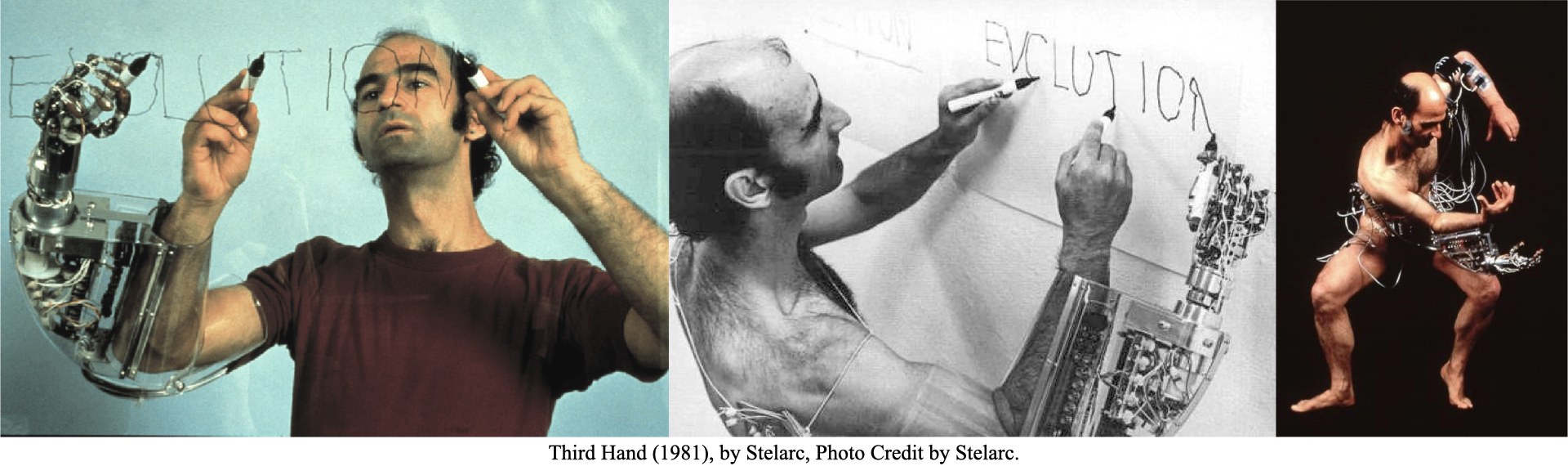}
\caption{The Third Hand by Stelarc.
\copyright Artist Mentioned.}
\Description{The Third Hand by Stelarc.}
\label{part1back}
\end{figure}

Firstly, cultural anthropology, focusing on the rich tapestry of human behavior and cultural expression, provides a unique lens for understanding and creating biological art. Artists challenge the existing cultural norms related to nature and technology, fostering a collaborative dialogue across disciplines. In 1981, Stelarc presented ``Third Hand (see Figure~\ref{part1back})," a performance that challenged the boundaries between the human body and technology by utilizing an articulated mechanical arm to write the word ``Evolution," addressing themes of information control and conditioned reflexes of bodily functions~\cite{mcleod2011stelarc}. He interrogated the nature of information control, but also the learned responses of the human body. This work exemplified the concept of a cyborg-like fusion, underscoring the symbiosis of flesh and machine. 

The work centered on the human being in biological art lies in its engagement with human interaction, control, and the body's extension through technology. Stelarc's piece directly integrates a technological appendage with the human body, challenging our understanding of the human form and its capabilities. Kac's work involves human participation to influence living organisms at a genetic level, reflecting on humans' deeper connection and control over biological processes. They use biological art to explore and critique the relationship between humans and technology, ultimately centering on human identity, evolution, and humans' intrinsic power in shaping and interacting with the biological world.

The emphasis of cultural anthropology focuses on people's day-to-day practices and cultural expressions and provides a rich and intellectually engaging framework for artists. It enhances their comprehension of biological concepts' diverse cultural interpretations and valuations. By challenging or reflecting on cultural norms concerning nature, artists investigate cultural biases and emphasize collaboration and cultural exchange. They can impact public opinion and policy by increasing awareness of ecological issues through exhibitions and public works. Art can function as a medium through which political concerns about migration issues, biological sciences, bioethics, environmental legislation, and the politics of health care can be deliberated. Biological art that celebrates genetic, species, and ecosystem diversity while advocating for a greater acceptance of various forms of life may be inspired by LGBTQ inclusiveness. It combines elements of these academic fields to produce culturally, politically significant, and biologically informed works. Their frequent objective is to stimulate intellectual engagement, question established beliefs, and incite discussions that transcend fields such as biology, culture, and politics.

Lévi-Strauss's structuralism, emphasizing the search for \textbf{universal patterns} in human thought and culture, originally came from a branch of anthropology and proposed to investigate cultural symbolism issues by drawing on this structuralist idea that all human societies, despite their apparent diversity, share specific cognitive frameworks and patterns of behavior~\cite{levi1963structural}. According to his theory, the human mind is a storehouse of many natural materials. It chooses elements in pairs that can be combined to create various structures. This methodology motivates artists to transcend the confines of social science by incorporating natural elements into their works to reveal and deconstruct the profound correlations between living organisms and the cultural environments that influence our nineteenth and twentieth break apart pairs of oppositions into individual elements to create new oppositions, including those between nature and culture, organic and synthetic, and innate and acquired.

Michel Foucault's biopolitical theory also influenced the artistic production of biological art~\cite{foucault2008birth}. To identify and analyze emergent logics of power in the nineteenth and twentieth centuries, Michel Foucault introduced the term \textbf{biopolitics} in his late 1970s lectures at the Collège de France. Foucault defines biopolitics as examining the mechanisms through which human existence assumed the status of a unique political concern within Western societies at the population level. It advocates for the importance of ``bios" (Greek for ``life") in all human pursuits, including government, science, technology, policy, education, art, and governance. It includes genetic and geographical variations among all life forms on Earth. Artistic expressions that challenge the manipulation and control of life forms—such as genetic engineering, immigration invasion, and the repercussions of political control on society—were influenced by his views on the governance of life and bodies by political powers.

In the book Gender Trouble (1990)~\cite{butler2002gender}, Judith Butler, a philosopher and gender theorist, initially employed the concept of \textbf{gender performativity}. She contends that one's biological sex does not dictate one's behavior. In contrast, individuals acquire the knowledge and skills necessary to conform to societal norms and expectations. The concept of gender can be understood as a deliberate enactment or display. Her ideas stimulated artists to challenge the dichotomous classification of sex and gender in biological contexts, fostering the production of art that investigates the malleability and enactment of identity. Butler's theory on gender performativity prompts the art world to reconsider traditional concepts of gender and biological determinism, advocating for a more flexible comprehension of identity. Artists are encouraged to produce artworks depicting the full range of biological and morphological diversity in nature, thereby capturing the intricate and diverse nature of gender expressions and identities observed in human societies.

\textbf{Tunnel Under the Atlantic by Maurice Benayoun}

\begin{figure}
\includegraphics[width=\textwidth]{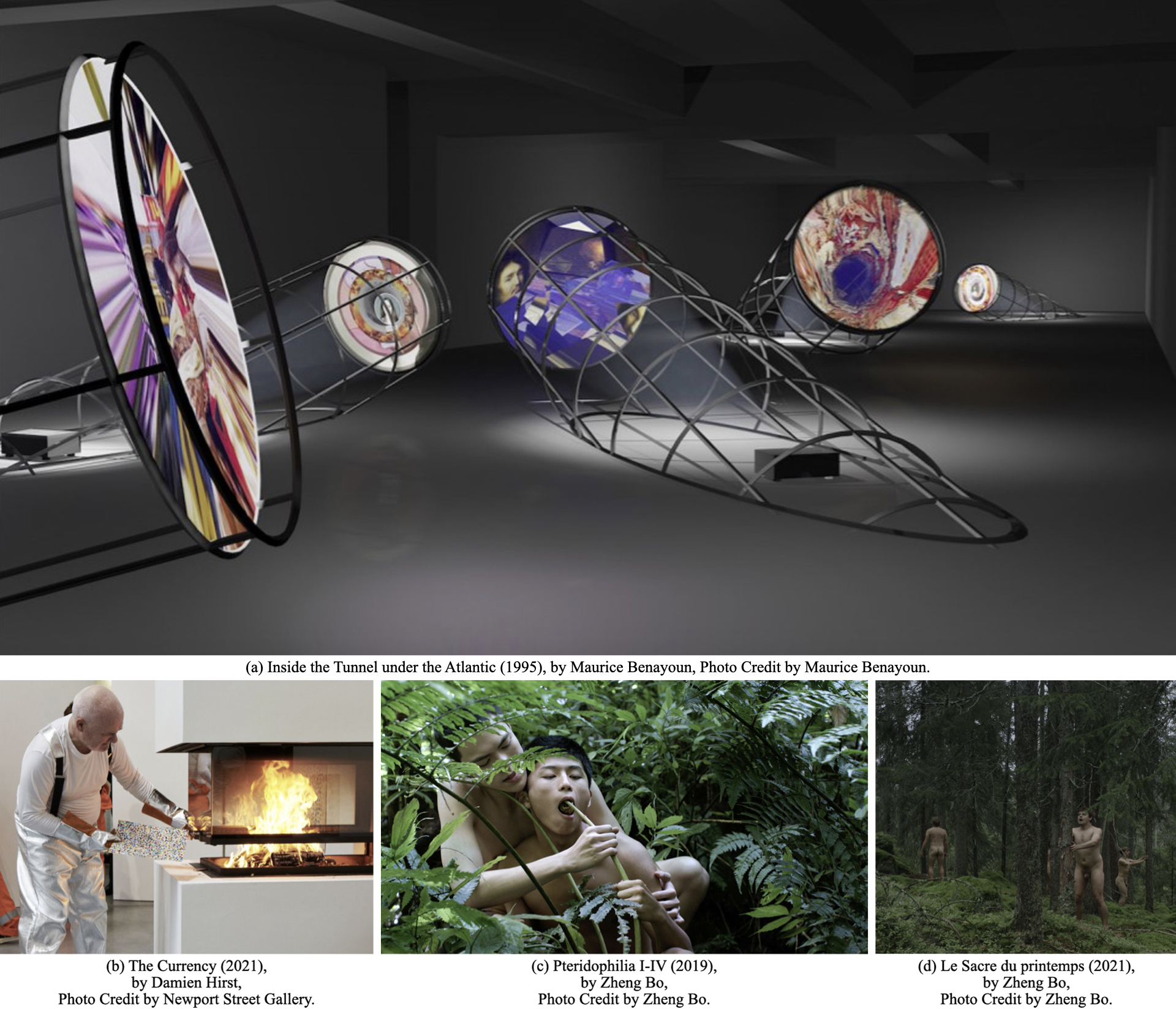}
\caption{The Cultural Anthropology Perspectives Concepts Reflected in the Artworks.
\copyright Artists Mentioned.}
\Description{The Cultural Anthropology Perspectives Concepts Reflected in the Artworks.}
\label{part1relate}
\end{figure}

Given the concept of cultural anthropology, many artists have been influenced to create a series of artworks in response to this idea. Maurice Benayoun developed an immersive virtual environment called ``Tunnel Under the Atlantic" in 1995~\cite{isea2016tunnel}. This project established a virtual connection between Paris and Montreal, enabling participants to explore a data-driven subterranean world (see Figure~\ref{part1relate}a). This virtual tunnel connects two distinct physical and cultural spaces. It reflects fundamental binary oppositions: Paris/Montreal, France/Canada, and Europe/North America, which formed ``cultural obstacles" and showcased the nascent digital opportunities. In this work,  participants interacted with cultural images used as ``obstacles," effectively decoding and reinterpreting them and enabling participants to actively interact with and modify cultural artifacts, highlighting their lasting significance.

Benayoun's work showcases the ongoing importance of universal patterns in comprehending the intricate nature of modern digital art and its connection to culture. It showed that art helps us understand and reconfigure cultural symbolic systems, and participants developed personal narratives as they navigated the virtual environment. This is consistent with Lévi-Strauss's view of myth as a structural element of culture in which individuals actively participate and contribute to the collective cultural narrative - the virtual tunnel became a metaphor for artistic creation. Meanwhile, drawing on Lévi-Strauss's concept of the bricoleur (savage mind)~\cite{mambrol2016claude}, this work recontextualizes available cultural materials to create new structures, demonstrating innovative use of existing cultural codes.

\textbf{The Currency by Damien Hirst}

Numerous valuable and irreplaceable art pieces have disappeared due to calamities, conflicts, or other force majeure events. Maintaining the original qualities of artworks over time is a significant challenge in the physical world. However, the advent of NFTs presents an alternative solution to this issue. While digital networks and identities also depend on non-permanent materials, they pivot the art world's focus from ``permanence" and ``originality" towards ``hybridity" and ``movement." In comparison, art preserved digitally in the Metaverse can outlast an artist's lifespan when adequately archived and maintained, all at a manageable cost. The unique and non-fungible nature of NFT tokens ensures that artworks can potentially endure as long as the internet.

In 2021, Damien Hirst introduced ``The Currency," which merges NFTs with tangible art, comprising 10,000 NFTs, each associated with one of 10,000 physical art pieces (see Figure~\ref{part1relate}b). Collectors of these NFTs face a choice: retain the digital token or trade it for the corresponding physical artwork~\cite{bbc2022hirst}. By offering collectors the choice between maintaining a digital token or exchanging it for physical artwork, Hirst sets up a political issue between the digital and physical ``bodies" of art. This choice underscores the tension between the traditional, institutionally managed life of art and the new, decentralized life of art in the digital realm. His work ensures their perpetual existence, and the transfer of ownership they offer deviates from traditional practices in the art market. The ``vitality" of the artwork is no longer tied to its material condition but to its existence within the digital ecosystem and the blockchain network. The ``hybridity" and ``movement" of art in the digital domain through NFTs emphasize the fluidity and changeability of value creation rather than the static notion of originality and permanence associated with physical art.

\textbf{Le Sacre du Printemps by Zheng Bo}

Zheng Bo's artistic oeuvre is deeply rooted in the biological gender performativity approach, where his engagement with flora extends beyond traditional mediums to encompass painting, dance, and film. ``Pteridophilia," a work by Zheng Bo, employs the sensual and intimate interaction between homosexual men and ferns to undermine conventional concepts of longing, sensuality, and affiliation (see Figure~\ref{part1relate}c). These challenges established limits and broadened the conversation surrounding sexuality and the natural world~\cite{labiennale2022bo}. It demonstrates that gender is enacted and understood through repetitive behaviors. Zheng challenges the dominant belief that humans are separate from nature by depicting a world where humans and plants have a close and sensual relationship. This portrayal of intimate interaction between humans and plants breaks the boundaries of traditional identity and social norms.

``Le Sacre du Printemps" (The Rite of Spring) at Tandvärkstallen is another remarkable example that goes beyond metaphorical representation to visually and actively depict the profound ecological connection between humans and plants through intricate dance (see Figure~\ref{part1relate}d). His works expanded the understanding of redefining the boundaries of human-plant relationships through stylized acts performed within a highly regulated framework.

\subsection{Plant Growth, Biosemiotics, and Ecological Aesthetics}

From cultural anthropological explorations of art, we see how technology and biology intertwine in expressing the human body and identity. Artists challenge our traditional understanding of nature and human identity through this fusion. Just as Stelarc's work reveals the boundaries between the human body and the mechanical, so do we begin to think about how this fusion affects our understanding of and interaction with the biological world. This reflection naturally leads us to the themes of Part II: plant growth, biosemiotics, and ecological aesthetics. In this shift, we move from a focus on the combination of human identity and technology to an exploration of the plant world and the environment. This is not only a shift in perspective from anthropocentric to ecocentric but also another aspect of the intersection of art and science. Suppose the first part had us thinking about how humans extend themselves and control biological processes through technology. In that case, the second part had us explore how humans interact with the wider ecosystem and discover the possibilities of beauty and harmony.

Plant growth represents the fundamental biological processes and the respective natural routine on the earth. However, the intricate dynamics of plant life encourage sustainable interactions between humans and the environment and provoke critical thought about the role of beauty and design in the natural world. The study of communication and signification in living organisms posits that plants engage in complex signaling behaviors with their species and their environment.

\begin{figure}
\includegraphics[width=\textwidth]{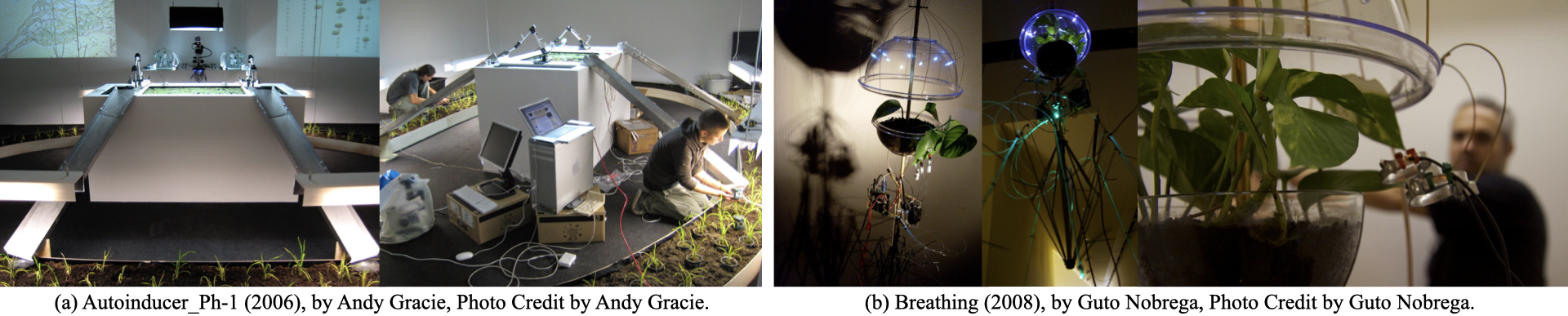}
\caption{The Plant Growth, Biosemiotics, and Ecological Aesthetics Concepts Reflected in the Early Artworks.
\copyright Artists Mentioned.}
\Description{The Plant Growth, Biosemiotics, and Ecological Aesthetics Concepts Reflected in the Early Artworks.}
\label{part2back}
\end{figure}

Andy Gracie's ``Autoinducer\_Ph-1" is a prime illustration of ecological aesthetics that merges traditional ecological practices with advanced technology~\cite{newexhibitions2006gracie}. This installation adopts Southeast Asian rice farming techniques, utilizing duckweed for natural nitrogen fertilization, spotlighting sustainable agriculture (see Figure~\ref{part2back}a). Gracie's work intricately interweaves living organisms with electronic and computational elements, creating a dynamic environment that explores the relationship between human technology and ecological systems. The work emphasizes the potential of harmonious coexistence between natural cycles, technological advancements, and plants' ecological relationships, suggesting that such synergy could be vital for the future of agricultural sustainability.

Guto Nobrega's ``Breathing" represents a more intimate scale of biological art, focusing on the individual experience of life and its interconnectedness~\cite{nobrega2008breathing}. This work is a compelling integration of a living plant with an artificial system, responding to its surroundings through an array of sensors and actuators, facilitated by an Arduino controller and a custom GSR circuit that translates biological impulses into movements and light patterns (see Figure~\ref{part2back}b). Integrating technology and biology prompts the audience to reflect on their position within the ecosystem, the processes by which living systems create meaning, and their interactions with the environment, presenting a microcosm of the more significant environmental issues at stake.

``Autoinducer\_Ph-1" and ``Breathing," both centered on plants, exemplify the duality of biological art – both an advocate for ecological harmony and a critique of the manipulation of life. They demonstrate that incorporating plants into works of art facilitates comprehension of environmental statements and stimulates a reassessment of the natural and artificial correlation. These artistic creations prioritize incorporating plants into their compositions, materializing ecological issues and encouraging viewers to reevaluate the relationship between advanced technology and organic existence.

The incorporation of a male urinal by Marcel Duchamp into an art exhibition as artwork fundamentally transformed the easel-painting-dominated Western art tradition. The action above gave rise to ``readymade object art," a form of artistic intervention that substantially broadened the parameters of what could be deemed creative. The artists portrayed vegetation on canvas, emphasizing the significance of plants rather than the artistic medium. They effectively converted the plant into a visual representation that assumed the central role in their piece.

On the other hand, the emergence of ecological art converted plants from passive subjects to active mediums. Plants emerged as narrative subjects within the artwork, communicating biological verities and eliciting emotional responses. This transition transformed the artistic medium by which practitioners shared their ideas and introduced the notion of ``deanthropocentrism" in art. They became vital elements of the art, enabling a conversation that moves beyond traditional portrayal to a comprehensive, ecological experience.

\textbf{Plant neurobiology} aims to unearth how plants perceive and act in an integrated and purposeful manner and how they do it~\cite{brenner2006plant}. It suggests that plants are capable of sophisticated behaviors such as sensing, learning, and communicating. It proposes an interdisciplinary and integrated view of plant signaling and adaptive behavior to study plant intelligence. The artists challenged our perception of plants as static or passive. By revealing the dynamic capabilities of plants, biological art invites audiences to reconsider the complexity of plant life and its agency. The interactive installations use real-time data from plants, such as electrical signals from plants' responses to stimuli, including touch, light, or gravity. Such signals are translated into visual or auditory information in color intensity and sound rhythm.

\textbf{Biosemiotics} was first used by Friedrich S. Rothschild and implemented by Thomas Sebeok, Thure von Uexküll, Jesper Hoffmeyer, and many others. It studies sign processes, or semiosis, in biological contexts~\cite{kull2007brief}. This theory posits that all living organisms, not just humans, participate in creating and interpreting signs. For plants, this refers to chemical signaling, growth patterns, and responses to environmental factors can all be seen as forms of communication. Artworks inspired by biosemiotics focus on the interpretative interactions between plants and their environment, demonstrating how plants communicate and represent a form of non-human intelligence. Art that visualizes the communication between plants makes these invisible processes perceivable to humans.

\textbf{Ecological aesthetics}~\cite{zeng2019construction} transcends traditional aesthetics by incorporating a reflection on and an evolution beyond the epistemological confines of ``traditional anthropology" and ``anthropocentrism." It challenges the narrow focus on artistic aesthetics that historically overlooks ecological contexts. This domain advocates for art that resonates with ecological tenets, emphasizing the interdependence of living entities and their collective meaning-making dynamics within the environment. Prompted by this theoretical framework, artists create works that depict or emulate the intricate ecological interplay between plants and their environment, encompassing relationships with other flora, fauna, and humans.

\begin{figure}
\includegraphics[width=\textwidth]{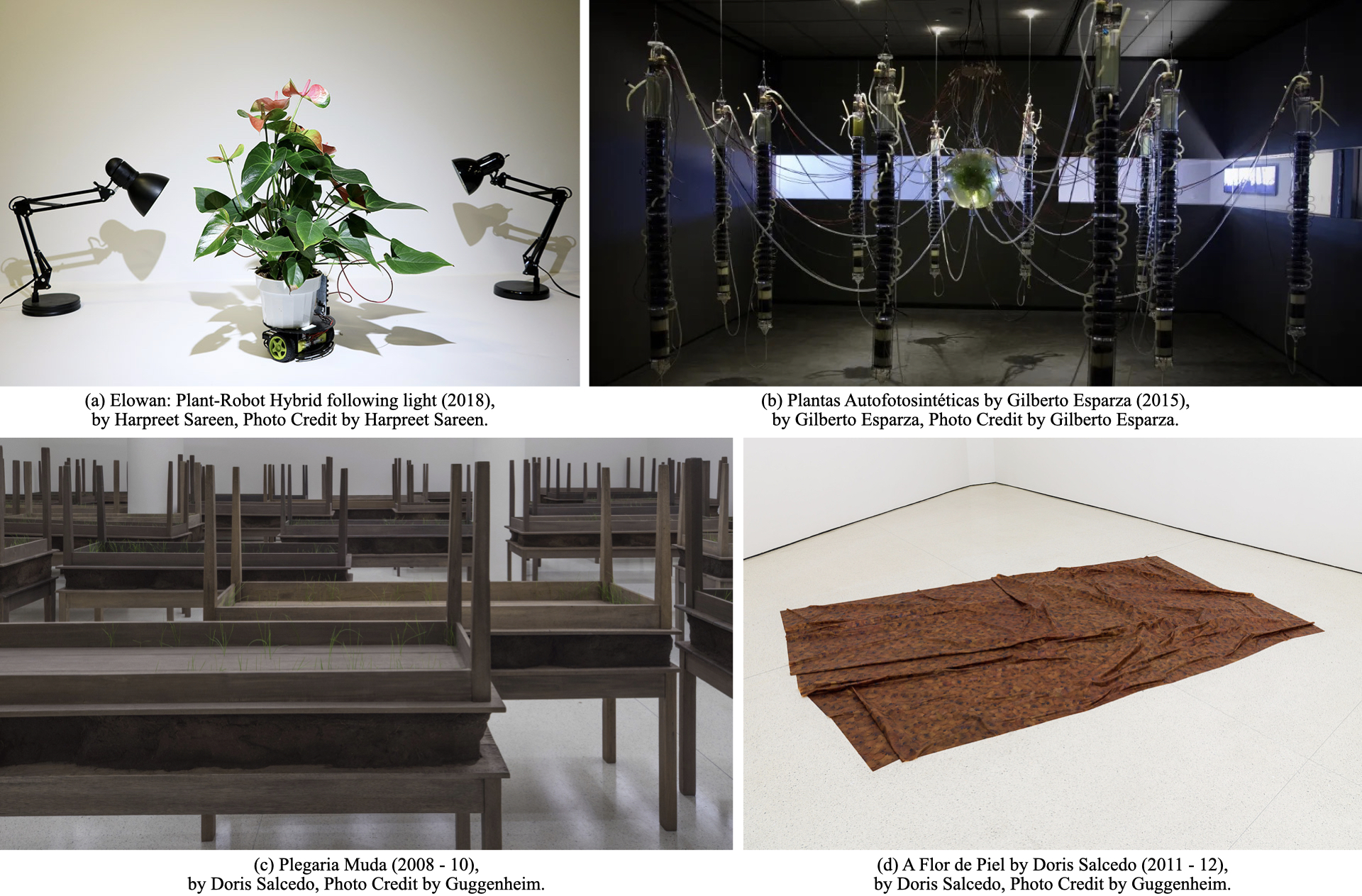}
\caption{The Plant Growth, Biosemiotics, and Ecological Aesthetics Concepts Reflected in the Artworks.
\copyright Artists Mentioned.}
\Description{The Plant Growth, Biosemiotics, and Ecological Aesthetics Concepts Reflected in the Artworks.}
\label{part2relate}
\end{figure}

\textbf{Elowan: Plant-Robot Hybrid following light by Harpreet Sareen}

Harpreet Sareen's work forges a new symbiotic model with plants. Elowan, he created, is a plant-robot hybrid featuring a wheeled robotic base outfitted with electrodes embedded in its leaves and stems (see Figure~\ref{part2relate}a). These electrodes intuitively respond to light and environmental stimuli by generating bioelectric signals. The detected signals trigger the robot's movement, guiding it toward or away from a light source. This technology translates and amplifies plants' reactions to external stimuli into observable physical movements. 

Elowan's actions are based on the plant's natural adaptation—modifying its growth, respiration, water absorption, and movement toward light sources. Unlike simple machines, these plant equipment can self-repair, self-power, and self-grow. This innovative symbiotic model represents a significant shift from an anthropocentric to a nature-centric approach, delving into the plant neurobiology of the plants' percept capabilities. As plants speak for themselves by expressing their growth needs, we harness technology to tweak their natural growth mechanisms. This work uses technology to change the natural growth mechanism of plants, allowing them to choose a living environment and state that is more suitable for them automatically.

\textbf{Plantas Autofotosintéticas by Gilberto Esparza}

Gilberto Esparza's works frequently delve into themes of nature, technology, and the interconnectivity of living systems. Plantas Autofotosintéticas (see Figure~\ref{part2relate}b) uses microbial fuel cells (MFCs) to harness the energy from photosynthesis in plants and translate it into sound, allowing for an auditory perception of a plant's living processes. When plants perform photosynthesis, they produce organic matter that bacteria in the soil degrade. As the bacteria metabolize these organic compounds, electrons are transferred to the anode of the MFCs, resulting in an electrical current. The current is then translated into sound, with fluctuations in electrical current resulting in non-speech audio. These resulting sounds provide an auditory perception of the plant's living processes, creating a direct sensory connection between the audience and the biological and ecological processes at work. This leads to a continuously evolving soundscape, which reflects the living state of the plants and their unseen interactions with the microbial world. 

Esparza's work invites the audience to listen to the usually inaudible and invisible energetic exchanges between plants and bacteria, allowing them to understand life's complexity. Plantas Autofotosintéticas visualizes plant communication, making invisible processes visible to humans and allowing all living things to contribute to the creation and interpretation of signs.

\textbf{Plegaria Muda and A Flor de Piel by Doris Salcedo}

Doris Salcedo's installations engage with themes of life, death, and the enduring cycles of nature. ``Plegaria Muda (Silent Prayer)" evokes this through its structure of wooden tables layered with earth, where grass sprouts signify resilience and the cyclical return to life amidst symbols of death (see Figure~\ref{part2relate}c). It consists of 83 tables paired with matching tables upended on them and natural grass growing in dirt set between the two tables' tops. This installation speaks to the ecological principle of life's tenacity and the environment's role in healing and renewal post-trauma. The grass, an element of the ecosystem, becomes a metaphor for the persistence of life and the possibility of regeneration after loss. 

In ``A Flor de Piel," Salcedo crafts a shroud from rose petals, a material choice that embodies the ecological concept of transience and the delicate balance of life (see Figure~\ref{part2relate}d). The installation reflects the vulnerability of emotion, which mirrors the organic decay and the temporal beauty inherent in ecological systems. The petals, sewn to resemble human skin, unfold stories of trauma, connecting the personal and the natural world in their shared fragility. Together, these works stand as poignant reminders of life's ephemeral nature while showcasing the capacity for beauty and healing within ecological processes, representing how ecological aesthetics reflect the balance of life. 

\subsection{Plants as Art Subjects and Mediums}

As we move from Part II, which focuses on integrating humans and technology and how they change our understanding of ourselves and the biological world, to Part III, we focus on the role of plants and ecological aesthetics. This shift is a shift in perspective from human-centered to ecocentric and another dimension of the intersection of art and science. How plants grow and communicate in nature offers a fresh perspective on the relationship between living things and their environment. In this transition, we begin to explore not only the role of plants in nature but also their dual role in art - both as depicted subjects and as mediums through which artists express themselves. This in-depth exploration continues the reflections on the interaction of humans, technology, and the natural world from Part II while also providing a new perspective on the unique place of plants in art and ecology. As these reflections deepen, this part will show the enduring appeal of plants in visual and intellectual culture and how they have been re-understood and re-shaped in works of art. This is a new understanding of plant biology and life itself and a rethinking of the relationship between humans and the natural world.

Plants have historically served dual roles in art as both subjects to be depicted and mediums through which artists express themselves, showcasing humanity's deep-rooted enchantment with the botanical realm. The meticulous botanical illustrations aimed at scientific documentation and the stylized interpretations pervasive across artistic movements highlight the persistent allure plants hold in our visual and intellectual cultures. This attraction is reflected in the observations of philosopher Emanuele Coccia in ``The Life of Plants: A Metaphysics of Mixture~\cite{coccia2019life}", where he posits that our understanding of biology, and indeed of ``life" itself, often neglecting the profound and distinct biological narratives of plants. 

Coccia's critique exposes an oversight within the life sciences and philosophy, where plants have been relegated to a lesser status than animals, primarily due to their perceived passivity and immobility. However, as Coccia illuminates, plants embody a form of continuous growth and development, unlike animals that typically halt physical development after reaching sexual maturity. This ever-expanding mode of existence allows plants to create new structures and organs across their lifespan, inextricably weaving their life processes with their environment. They claim space not through locomotion but through expansive growth and intricate interactions with their surroundings to procure nutrients and sustain their livelihoods. 

So, the plants' nature of expansion encourages artists to transcend traditional representations of plants, integrating plants into their work and using the invasion of nature to expand the lines between art and biology. As both subjects and mediums, plants challenge artists and audiences alike to reconsider the boundaries between the animate and inanimate, the made and the grown, and to appreciate the inherent creativity found within the natural world and displayed in ways that prompt reflection on ecological processes, environmental concerns, and the interconnectivity of life.

Navigating the vast continuum between order and the unavoidable descent into disorder, biological art emerges as a profound commentary on the nature of life itself. With its living canvases and organic mediums, biological art provides a fresh perspective on the age-old human quest to comprehend and influence the natural world. As these works develop, evolve, and occasionally deviate from their intended paths, they become metaphors for the more prominent themes of life's unpredictability, the difficulty of maintaining control over nature, and biological systems' remarkable adaptability. Just as humans have learned to navigate and transform diverse landscapes, biological art captures the essence of this adaptable spirit. Often dynamic and ever-changing, these artworks testify to the enduring relationship between life and its surroundings.

\textbf{Cybernetics}, proposed by Norbert Wiener, as a study of regulatory systems, feedback, and control within machines and living organisms, can take on a new dimension when systems go out of control~\cite{wiener2019cybernetics}. The biological processes do not behave as expected because they have natures and are always affected by their surroundings, leading to unanticipated outcomes. This reflects themes such as the unpredictability of life, the limits of human control over nature, the adaptability of biological systems, and understanding how humans adapt to the dynamic, changeable environment. Artists take inspiration from the binary concept of control and chaos, the balance between the artist's control (cybernetics) and the inherent chaos within living systems when they diverge from the expected path.

John Stuart Mill's seminal work, \textbf{On Liberty}, advocates for individual freedom and explores the boundaries of society's control over individuals~\cite{mill1966liberty}. When plant and biological art goes out of control, it can evoke questions about the liberty of the living components within the artwork. Such occurrences question the extent of the creator's and society’s dominion over life, underscoring the need to honor the innate autonomy and the unpredictable nature of living systems. When biological elements in art diverge from human expectations, it foregrounds a debate on the moral implications of our interventions in natural life cycles. The concept of 'freedom' for these entities from human constraints becomes a poignant critique of our ethical responsibilities in manipulating biological processes.

\textbf{Adaptationism}, a concept introduced by the British geographer P.M. Sauer in 1930, referred to the idea that many traits in organisms have evolved through natural selection because they offer survival and reproductive advantages~\cite{watts2015origins}. This viewpoint underscores the role of adaptive traits in enabling organisms to adjust to their environments, with these traits being honed over time by natural selection. It has evolved into a doctrine in human-environmental relations, acknowledging the dynamic interplay between the natural environment and human activities. This term captures the balance between the constraints imposed by the natural environment on human endeavors and the capacity of human societies to exploit and transform the natural environment to suit their needs. Human beings must adapt to new environments for development, such as the arid conditions of Phoenix, Arizona, or the dense urban landscapes of Tokyo, Japan. Inspired artists reflect these scenarios from their biological artworks.

\begin{figure}
\includegraphics[width=\textwidth]{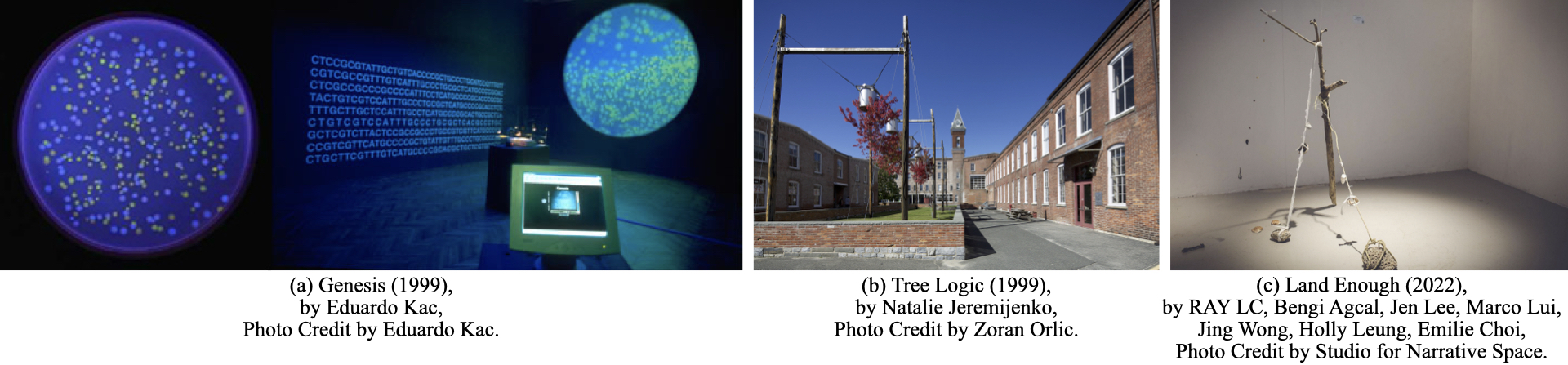}
\caption{The Plant as Art Subjects and Mediums Concepts Reflected in the Artworks.
\copyright Artists Mentioned.}
\Description{The Plant as Art Subjects and Mediums Concepts Reflected in the Artworks.}
\label{part3relate}
\end{figure}

\textbf{Genesis by Eduardo Kac}

Eduardo Kac's ``Genesis" discusses the unpredictable territory where art meets biotechnology (see Figure~\ref{part3relate}a). Kac transcribed a biblical excerpt into Morse code and subsequently converted this sequence into DNA base pairs, which were then inserted into living bacteria. The installation in a gallery included an interactive element where the audience could initiate mutations within the bacterial DNA by exposing it to ultraviolet light~\cite{kac1999genesis}. This process altered the original Morse code and, symbolically, the biblical narrative itself, thus underscoring the theme of losing control. 

Through this living artwork, Kac poignantly illustrates the unanticipated outcomes inherent in biological systems and raises questions about the extent and ethics of human manipulation of life at the genetic level. The project, intersecting with the principles of cybernetics, becomes a commentary on the chaos that follows when organic processes are subject to technological intervention and control.

\textbf{Tree Logic by Natalie Jeremijenko}

``Tree Logic," as an installation, subverts the expected growth patterns of trees by planting them upside-down (see Figure~\ref{part3relate}b). The essence of the artwork doesn't lie in a singular moment but rather in the transformation of the trees over time, showcasing their dynamic nature. This installation challenges natural tendencies by inverting the trees, which continue to grow toward the sun, potentially creating highly unusual forms over time. This inversion prompts viewers to reconsider the nature of what is natural by observing trees not as static symbols but as active, responsive organisms. 

The work serves as a living demonstration of the tension between intrinsic biological impulses and externally imposed conditions. In ``Tree Logic," Natalie Jeremijenko highlights the contrast between the public's interaction with art and science. It underscores the importance of respecting the inherent autonomy and the unpredictable evolution of living systems.

\textbf{Land Enough by RAY LC, Bengi Agcal, Jen Lee, Marco Lui, Jing Wong, Holly Leung, Emilie Choi}

The ``Land Enough (see Figure~\ref{part3relate}c)" pop-up exhibition, featuring interactive design fiction artworks, aims to catalyze climate action~\cite{recsf2022enough}. Curated by Narrative Space Studio, with key figures like Ray LC and local researchers, the exhibition fosters environmental awareness through four distinct teams: Blue Team constructs protective beachfront shelters against tidal flooding; Red Team creates maps to identify nuclear and biological waste sites; Green Team builds a community-engaging renewable energy generator; and Yellow Team curates artworks that educate and uphold traditions, illustrating ecological changes and connections. Each group's unique approach underscores the collective effort needed to address the pressing challenges of climate change.

Each team's contribution addresses broader environmental concerns and emphasizes the role of plants and vegetation in mitigating climate change. ``Land Enough" becomes a microcosm for plant-centered ecological thought, where every initiative contributes to a greater understanding and appreciation of plants' central role in maintaining environmental balance and combating climate change. These efforts emphasize human reactions toward ecological changes.

\subsection{Ecological Art in the Metaverse}

As our understanding of plants as vivid mediums and creative subjects in art has deepened, we have begun to look at how art adapts to and reflects rapidly changing ecological dynamics and technological advances. In this environment, artists are exploring traditional themes in nature and seeking expression within the new digital realm. This exploration leads us to the second part of the discussion. This part will focus on how ecological art navigates the tide of digital transformation, especially as it unfolds in virtual spaces such as the Metaverse. This shift is not only a renewal of artistic expression but also a reconsideration of ecological impact, in which artists use digital platforms to mitigate the environmental impacts that physical art production can have. Thus, as we explore this section, we will see how artists create in the virtual world while maintaining a focus on ecological themes and sustainability. This represents a new form of interactive relationship between art and nature and a new exploration of art in terms of ecological awareness and technological innovation.

Amid the rapid circulation of capital, there has been a significant shift in ecological dynamics and individuals' reliance on silicon-based technologies. In response to environmental challenges, ecological art has begun to navigate the digital landscape, increasingly finding a home within the virtual expanse of the Metaverse. This shift allows artists to create within a new realm, one where the environmental impact of art can be mitigated by considering the production and energy demands of the necessary hardware. By operating in the Metaverse, artists can circumvent some environmental costs associated with physical art production while exploring the constraints imposed by the tangible world through a novel digital framework.

This shift can also be understood as a case of what Lev Manovich proposes as ``Info-Aesthetics"~\cite{negri2009antinomies}. Although both biological art and digital art can be regarded as parallel practices of circumventing anthropocentric-ism, especially from Hayles' perspective of ``Posthuman"~\cite{hayles2000we} transcendence, it was not until this epochal transformation from modernity to ``information" with postmodernity as a transitory phase surfaces to exert visible impact on society that the intertwinement between biological art and digital art becomes reasonable and critical. The emergence of ``Info-Aesthetics" makes the plant-centered way of social organization a double-reference for navigating digital activities as well, while how humans process information also affects the forms and evaluation of biological art. This intertwinement provides the foundation for our case studies and data analyses.

\begin{figure}
\includegraphics[width=\textwidth]{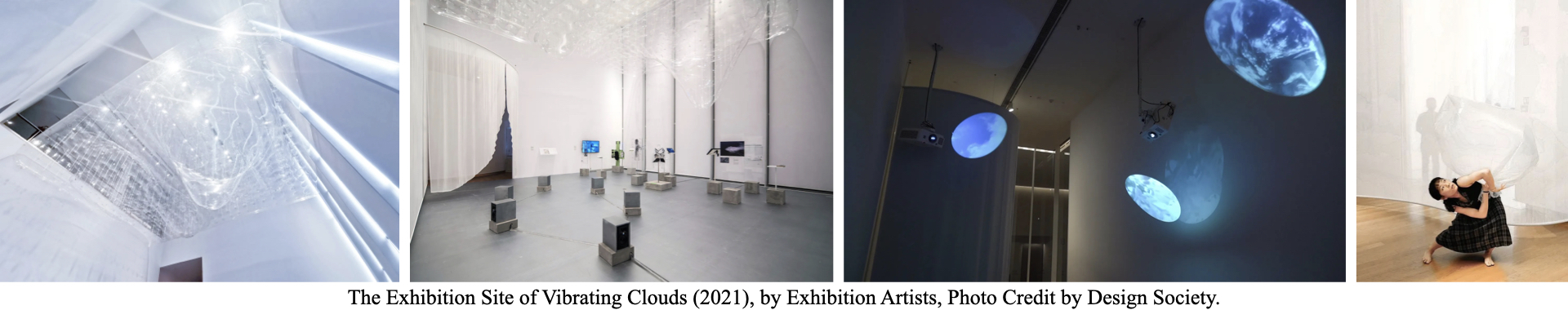}
\caption{The Ecological Art in the Metaverse Concepts Reflected in an Exhibition.
\copyright Artists Mentioned.}
\Description{The Ecological Art in the Metaverse Concepts Reflected in an Exhibition.}
\label{part4back}
\end{figure}

The exhibition ``Vibrating Clouds," inaugurated in 2021 at the Sea World Culture and Art Center's Garden View Gallery in Shekou, China, exemplifies this transition (see Figure~\ref{part4back}). Curated by Cai Yixuan, a Design Council Curatorial Award recipient, the exhibition featured contributions from twelve scholars from diverse international backgrounds and disciplines~\cite{conversazione2022clouds}. The exhibition presented twelve pieces of work intersecting meteorological study, aesthetic exploration, and architectural innovation. The participating artists employed the concept of air to reflect on the essence of life, intimacy, and freedom. Their works also offered a poignant commentary on the contemporary global condition, contrasting air as a symbol of dynamic movement with the restrictive nature of pandemic-induced controls, social distancing, global protectionism, and the proliferation of digital surveillance. Visitors to ``Vibrating Clouds" were invited to engage with the pre-modern notions of ``breath" and ``sky" through a multisensory experience involving visual, auditory, and tactile elements. In doing so, the exhibition facilitated an ecological engagement within a sustainable context, pushing the boundaries of what constitutes immersive and nature-centric art. Through this lens, ``Vibrating Clouds" serves as a testament to the potential of ecological art to harmonize sustainability with the expanding frontiers of artistic immersion.

\begin{figure}
\includegraphics[width=\textwidth]{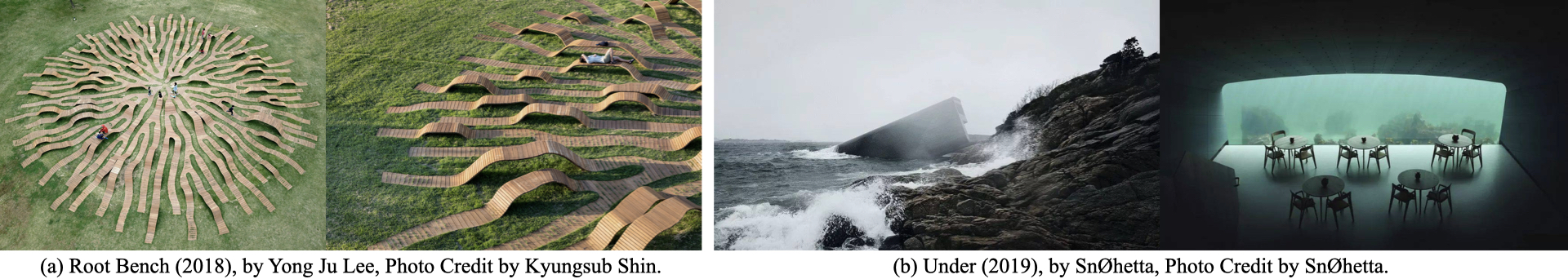}
\caption{The Natural Environment Concepts Reflected in Architecture and Landscape Architecture.
\copyright Artists Mentioned.}
\Description{The Natural Environment Concepts Reflected in Architecture and Landscape Architecture.}
\label{part4theory}
\end{figure}

Certain philosophies deeply ingrained in the interplay between urban landscapes and human endeavors have guided the architectural world. Many architectural projects have been conceived by drawing inspiration from the natural environment and blending man-made structures with flora to cultivate novel living experiences for humans. An example is the ``Root Bench," a sprawling circular installation by Yong Ju Lee, which seamlessly merges functionality with organic form and guides the human activities of relaxing in nature (see Figure~\ref{part4theory}a). Another groundbreaking example is Europe's first underwater restaurant, ``Under," a visionary project by Snøhetta, which immerses diners in an aquatic dining experience, marrying the built environment with the marine ecosystem (see Figure~\ref{part4theory}b). These projects exemplify how architecture can harmonize artificial elements with nature to redefine and re-guide human activities.

The concept of \textbf{sustainability} has its roots in the environmental movement of the 1960s and 1970s. Considerations include energy efficiency, resource conservation, environmental protection, and social responsibility. This sustainable design aims to reduce buildings' environmental impact by increasing efficiency and moderation in the use of materials, energy, and the ecosystem as a whole. However, as a continuous circular motion, it also represents the plants' reaction forces on humans. New plants growing according to the structure, such as ``Root Bench," alter how people gather and unwind.

Renowned architects like Frank Lloyd Wright and Louis Sullivan have been acclaimed for their design philosophy that advocates for organic architecture, one that considers the natural surroundings and seeks to be in harmony with them and reflects a concept of \textbf{digital continuation of techno-organic}.

In the context of environmental sensitivity and community engagement, the design process considers the immediate environment and its vegetation. It allows artists and designers to interact with the local ecosystem. Furthermore, contemporary initiatives leverage digital technology to promote ecological consciousness. For instance, movements like ``Internet + National Compulsory Tree Planting" have emerged, using digital platforms to encourage environmental stewardship. A noteworthy example is integrating the Alipay Ant Forest tree planting model into the national voluntary tree planting system~\cite{xiong2018analysis}. This model rewards users with virtual ``green energy" points for making eco-friendly choices, which can be used to plant real trees, thereby gamifying and incentivizing reforestation efforts. Such innovative approaches represent a fusion of digital engagement and environmental action, reflecting the growing trend of using technology to foster sustainable practices.

\textbf{Giving Tree by HUSH}

\begin{figure}
\includegraphics[width=\textwidth]{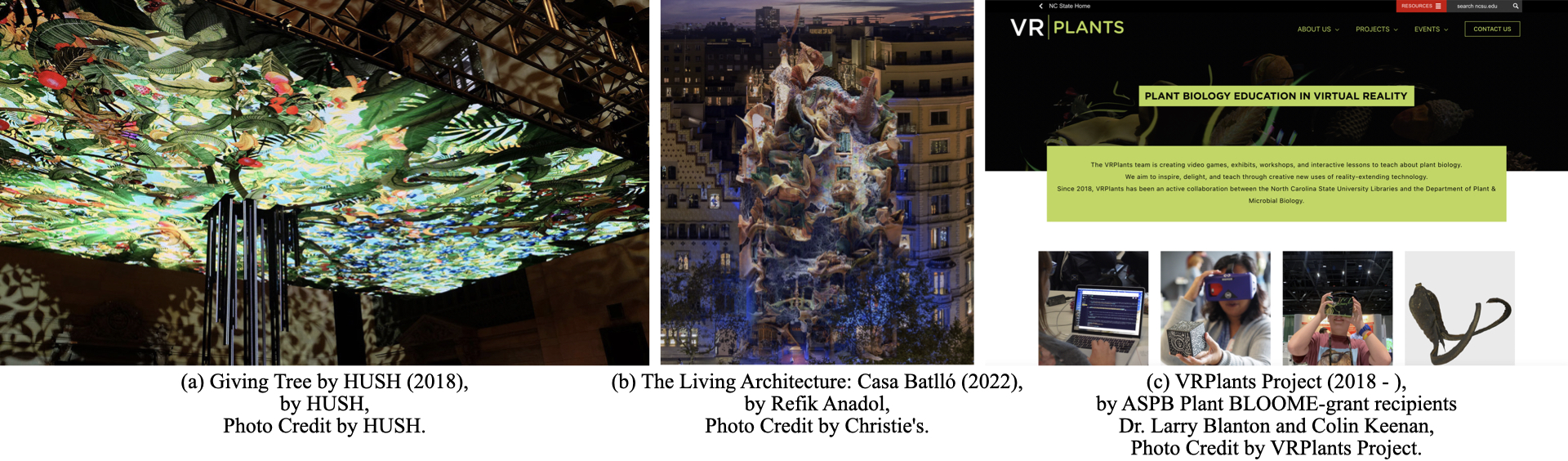}
\caption{The Ecological Art in the Metaverse Concepts Reflected in Artworks.
\copyright Artists Mentioned.}
\Description{The Ecological Art in the Metaverse Concepts Reflected in Artworks.}
\label{part4relate}
\end{figure}

The Giving Tree installation at Grand Central Station is an interactive technology for celebrating Chobani's 10th anniversary (see Figure~\ref{part4relate}a). Upon entering the venerable space of Grand Central, visitors are greeted by the unexpected sight of a modernist gesture set against the backdrop of historic architecture. The installation, a collaboration with the HUSH team, invites participants to engage in the act of planting digital seeds. This interactive experience is visually captivating and purpose-driven, with each digital seed resulting in a donation of Chobani yogurt to No Kid Hungry, a campaign dedicated to ending childhood hunger in America. As visitors embrace the glowing orbs at the structure's base, their touch breathes life into a sprawling digital canopy displayed on an overhead screen. The touching symbolizes the warmth of human generosity, which manifests as flourishing virtual branches and leaves. The metaphor is powerful because individual actions can contribute to a collective force of growth and nourishment. Commuters, often in haste, are given pause to contribute to a larger cause, merging the rhythm of daily life with the rhythm of giving. It led to the gathering of activities of passengers and encouraged them to be involved in being aware of childhood hunger. The Giving Tree is a testament to the potential of biological art to react to human behavior to form such gathered activities and the power of plants in any form, both physical and virtual.

\textbf{Living Architecture: Casa Batlló} 

The dynamic NFT created by Anadol is more than just a digital representation of Casa Batlló's façade. It's a living, breathing entity that exists in the digital ecosystem, reflecting the complexities and essence of the original building in Barcelona~\cite{christies2023}. This project goes beyond static replication by incorporating human interaction and environmental changes, similar to how a biological entity responds to its surroundings (see Figure~\ref{part4relate}b). Anadol's machine-learning algorithms allow the digital twin to evolve, mirroring the constant transformation inherent in biological art. Like a living organism, Casa Batlló's digital façade grows and changes. By incorporating data that captures human experiences and environmental variables, the artwork becomes an intersection where the physical and digital worlds collide, engaging viewers on a deeper level. The ``The Living Architecture: Casa Batlló" NFT exemplifies the potential for digital twins to preserve and perpetuate cultural heritage. In a world where the original is susceptible to decay or destruction, digital twins provide a form of immortality, preserving the essence of iconic structures.

\textbf{The VRPlants Project by ASPB Plant BLOOME-grant recipients Dr. Larry Blanton and Colin Keenan}

The VRPlants project is an innovative and immersive educational platform that harnesses the power of virtual reality (VR) to teach plant biology and the significance of environmental conservation (see Figure~\ref{part4relate}c). Led by Dr. Larry Blanton and Colin Keenan, recipients of the ASPB Plant BLOOME grant, the project offers a compelling suite of VR experiences designed to educate and inspire users about the intricate world of plants and their pivotal role in our ecosystem. It offers three forms of Form and Function: Ram’s Horn,  The Naturalist’s Workshop, and VR Field Trip: Longleaf Pine Restoration.

``Ram’s Horn," a web-based virtual reality game, offers learning through play and deepens the understanding of plant form and function relationships. Players take on the role of the Proboscidea Louisiana, navigating the challenges of seed dispersal in a simulated environment. The Naturalist’s Workshop expands the educational potential of VR by providing a virtual hands-on experience with botanical specimens and naturalist tools. Designed for the Oculus Quest, this application is a brilliant example of how museums can incorporate modern technology to enhance the visitor experience. The 360-degree videos documenting the restoration of the longleaf pine ecosystem are a testament to the project's dedication to connecting users with real-world conservation efforts. These virtual field trips allow students to explore and learn about restoration ecology safely and sustainably. The controlled burn and its role in supporting longleaf pine growth are showcased in an educational and profoundly impactful format, highlighting the delicate balance and necessary interventions required in ecosystem management.

This hands-on approach to learning encourages users to think critically about plant adaptations and survival strategies in a way that a traditional classroom setting may not. By simulating a naturalist's workspace, users are offered an intimate and detailed exploration of the botanical world, which can ignite a passion for natural sciences and underline the importance of preserving such knowledge. It also reflects the novel method of biological art application in the education community.

\section{Methodology}

We proposed a novel theoretical framework for analyzing these reviewed contents to address the research questions adequately. We also combined the mixed-methods approach, such as artwork and medium analysis, quantitative analysis, and case study examinations.

\subsection{Theoretical Framework ``Biocentric-Creation Transformation Ideology"}

The BCTI (see Figure~\ref{gao}) we proposed represents a paradigm shift in the perception and purpose of art, moving away from an anthropocentric (human-centered) viewpoint towards an ecocentric (ecology-centered) perspective and then to a perspective of ecology-reacted-human activity. This ideology underscores the belief that art should not solely manifest human creativity and experience but should also acknowledge and integrate the intrinsic value of non-human life forms and their ecological functions.

In this framework, human creativity is seen as the initiator rather than the culmination of the artistic process. Artists begin by cultivating plants or incorporating living biological systems into their art, setting the stage for a continuous and often unpredictable transformation led by biological entities. As these plants grow and proliferate, they do so beyond the strict control and foresight of the artist, in turn influencing human experiences and perceptions of the artwork. This transformational ideology suggests that plants' dynamic growth and behavior become a part of the artwork's evolution, creating a living art piece that changes over time. As plants engage in natural processes, they can alter the artwork's aesthetic, structural, and conceptual elements. They reflect the ecological processes and cycles essential to life on Earth.

The BCTI posits that plants' collective existence and vitality are pivotal in impacting human communities by fostering an ecological consciousness and a deeper connection with the natural world. Therefore, Biological art becomes a medium through which ecological relationships are represented and actively embodied and engaged, promoting a more harmonious coexistence between human and non-human actors. By challenging traditional boundaries and extending the utility of art to encompass broader ecological relationships, this approach aims to reshape the discourse around art and its role in society. It advocates for a more inclusive understanding of aesthetics that recognizes the interconnectedness of all life forms and their shared environments, emphasizing the importance of ecological balance and sustainability in artistic expression.

\subsection{Analysis of Artworks and Mediums}

The study curated a selection of biological artworks focusing on plant-centered art and mediums that reflect our transformation stages of ideologies, including using living plants or organic materials and incorporating technology or interactive methods that challenge traditional art forms. The methodology for analyzing the artworks ranges from content and historical analysis, iconographic analysis, and interdisciplinary analysis from sociological, art history, science, and biological aesthetics.

Content and historical analysis focus on the messages, subjects, and symbols within artworks, examining the information the artist intends to convey and the meanings that viewers might derive from the work~\cite{afrough2023content}. This approach considers how the time period and cultural environment influence the creation and interpretation of art by placing the artwork within its historical and cultural context, indicating how the time's social, political, economic, and artistic trends influenced the artwork's creation. This study draws on cultural anthropology to analyze art creation and relates the artwork to structuralism, biopolitical theory, and gender performativity. By analyzing the artworks' visual salience preferences, such as motifs, themes, and representations of images, bodies, and shapes, we place the artwork within the frameworks of structuralism, biopolitical theory, and gender performativity. These contextual analyses enrich our understanding of art as a social artifact created within complex networks of meaning and power. The emphasis of structuralism on the relationships between elements within a culture helps us reveal how art functions as part of a broader cultural code. Art can reflect and critique how societies exert control over bodies, often highlighting issues of power, sovereignty, and the body politic. Furthermore, art serves as a medium through which the complexities of identity, expression, and societal roles are negotiated and represented. These theoretical frameworks provide a foundation for situating the artwork within a broader sociopolitical and cultural discourse.

The iconographic analysis studies the symbolism and allegory in the work, along with their traditional and cultural meanings. Erwin Panofsky proposed a three-level framework for interpreting artworks: primary or natural subject matter, secondary or conventional subject matter (iconography), and intrinsic meaning or content (iconology) to explore how the artist uses these images to convey deeper meanings~\cite{panofsky2018studies}. The methodology is rooted in understanding how visual symbols and their arrangements communicate layers of meaning that transcend their immediate appearance. It is pivotal for decoding the latent content that artists embed within their works. From the previous content and historical analysis, we clarified the visual saliences of the artworks and emphasized their combination. By connecting the artwork's visual saliences with iconographic interpretation, we reveal how biological artists communicate profound messages about society, identity, and the human condition through their careful choice of imagery and thematic focus. The iconographic analysis thus serves as a crucial tool for unraveling the intricate web of meanings within an artwork, offering a richer understanding of its place and purpose within the tapestry of cultural expression.

And the interdisciplinary analysis, including the aspects from sociology, art history, science, and biological aesthetics that embrace the application of various disciplinary perspectives to understand artworks~\cite{law2010interdisciplinary}. The interdisciplinary analysis that integrates aspects from sociological, art history, science, and biological aesthetics offers a comprehensive framework for understanding and interpreting artworks. This approach allows us to consider the societal context within which art is produced and received, acknowledging the influence of cultural, economic, and political factors. Art history contributes a temporal and stylistic dimension, tracing the evolution of artistic trends and their historical significance. Scientific perspectives, especially from biology and plant neurobiology, provide insights into the living components of art and the dynamic interactions between organic and inorganic elements. Biological aesthetics focuses on the intrinsic beauty of natural processes and forms and their representation or incorporation into art. Together, these diverse perspectives enrich our analysis, enabling a more nuanced appreciation of the complex ways art intersects with human experience and the natural environment.

\subsection{Case studies}

In the digital field, specifically within the Metaverse, ecological art assumes new forms and meanings, necessitating a novel approach to case study analysis~\cite{allam2022metaverse}. By conducting case studies on ecological art within digital platforms, this study investigates how these immersive environments redefine engagement with ecological aesthetics and concepts. The Metaverse provides a platform for examining how virtual representations of plants and ecosystems challenge our perceptions of space, presence, and materiality in art. It offers a unique opportunity to explore how artists address ecological concerns in a rapidly evolving digital landscape, potentially transforming viewers' perceptions and interactions with the natural world. Through such studies, we critically assess the potential for ecological art in the Metaverse to influence real-world attitudes toward the environment and plant life, highlighting art's evolving role in technological advancement and environmental consciousness.

Our case study methodology, with its in-depth focus on transformation ideology, is well-suited for exploring the intersection of biological art with natural sciences and within the Metaverse. Our study uncovers how plants' biological and ecological significance is interpreted and represented across various cultural contexts by examining specific instances where plant life is central to artistic expression. This exploration delves into how artists incorporate live plant growth into their works as an ideology, with shaped forms and environmental reflections, and what plants signify within different cultural mythologies through the lens of interdisciplinary perspectives. Furthermore, we investigate how biosemiotics—the communication, representation, and effect of plants within their ecosystems—influence and are depicted in artistic creation. Our approach reveals the complex dialogues between human cultural practices and the non-human natural world, offering rich insights into how plants are not merely subjects but also active mediums that shape the artistic narrative.

\section{Results}

We conducted a review of biological art works archived in three major platforms for media art---Ars Electronica\footnote{\url{https://ars.electronica.art/news/en/}}, ISEA\footnote{\url{https://www.isea-archives.org/}}, and ADA\footnote{\url{https://digitalartarchive.at/}}. These archives were chosen due to their prominence as leading repositories of new media art. However, some other entities, like CARTA\footnote{\url{https://carta.archive-it.org/}}, Rhizome\footnote{\url{https://rhizome.org/}}, Rhizome ArtBase\footnote{\url{https://artbase.rhizome.org/wiki/Main\_Page}}, oldweb.org\footnote{\url{https://oldweb.today/\#19960101/http://geocities.com/}}, Art Daily\footnote{\url{https://artdaily.com/}}, Artlinkart\footnote{\url{http://www.artlinkart.com/cn/}}, etc., were not included in our analysis due to limitations in database capacity and media compatibility across different archives. We leveraged the genre classifications used on the ADA website to categorize relevant biological artworks. This categorization allowed us to identify 117 biological artworks in the ADA archive dating from 1979 to the present. Due to the archive's 5-year categorization cycle, the works were distributed as follows: 2021--present, 4 works; 2016--2020, 15 works; 2011--2015, 23 works; 2006--2010, 31 works; 2001--2005, 24 works; 1996--2000, 18 works; 1990--1995, 12 works; 1979--1989, 1 work. A key limitation of the ADA archive is the lack of a peer-review or screening system for work submissions, unlike curated archives like Ars Electronica and ISEA, which have organizing committee reviews. This means the quality and relevance of all works in ADA cannot be ensured. However, we included ADA in our analysis to provide a more comprehensive view of biological art over time while noting its limitations. Further analysis only focused on works from the rigorously curated Ars Electronica and ISEA archives to ensure relevance and accuracy when discussing specific biological art examples and trends. Expanding this curated archive analysis with broader datasets remains an area for future work.

We analyzed award-winning biological art projects from two leading media art archives - Ars Electronica and ISEA - to identify trends over time. 

At Ars Electronica, we reviewed Grand Prix winners over 2016-2023. Of 16 total winners, ten centered on biological art themes. In 2016-2019, 3 out of 8 (37.5\%) were biological art-focused. Strikingly, this rose to 7 out of 8 (87.5\%) for 2020-2023, indicating a significant increase in bio-art recognition.

In the ISEA archive, 52 biological artworks spanned 1993-2020, categorized by year presented rather than creation year. The years 2013 and 2020 stood out, prompting a comparative study. This revealed several shifts between these years, including:

\begin{itemize}

\item  Greater focus on environmental and climate themes in 2020 versus more individualistic concerns in 2013.

\item More interactive and immersive pieces in 2020, with 2013 emphasizing static installations. 

\item More works incorporating AI and data in 2020 compared to traditional media like sculpture in 2013.

\end{itemize}

Our analysis of these prestigious archives demonstrates a growing mainstream appreciation of biological art, with recent years showing particular spikes in recognition. Comparing specific years also illuminates evolving biological art forms and concerns. Further analysis of additional archives and years would enrich perspectives on biological art progression.

\subsection{Data Analysis on Ars Electronica}

\begin{figure}
\includegraphics[width=\textwidth]{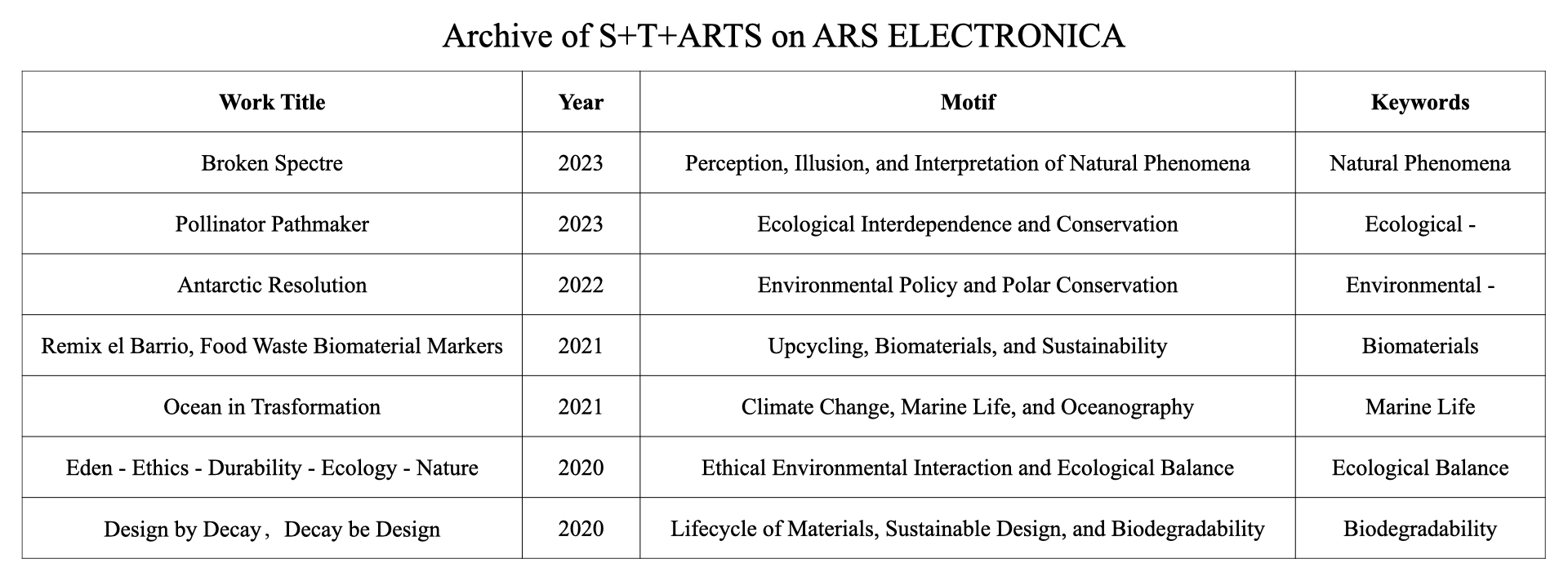}
\caption{The Projects of Grand Prizes on the archive of S+T+ARTS on ARS ELECTRONICA from 2020 to 2023.
\copyright Table Credit by Authors.}
\Description{The Projects of Grand Prizes on the archive of S+T+ARTS on ARS ELECTRONICA from 2020 to 2023.}
\label{archive1}
\end{figure}

Drawing from the S+T+ARTS archives at ARS ELECTRONICA, we delve into the grand prize projects related to biological art over the last four years, spanning from 2020 to 2023 (see Figure~\ref{archive1}). The STARTS Prize is a beacon of achievement, celebrating the pinnacle of creative ingenuity where science, technology, and the arts converge. Two exceptional projects are distinguished yearly with grand prizes, acknowledging their singular contribution to this interdisciplinary nexus.

Among the eight celebrated works recognized by the STARTS, seven are notable for their contributions to the field of biological art. This genre is marked by the intersection of biotechnology, scientific thought, and artistic expression, where living processes and biological systems become a medium and subject for creativity. 

Compared to the period from 2016 to 2019, 2020 represented a significant uptick in selecting works centered on biological themes for the STARTS Prize. This increase indicates an expanding trend toward melding life sciences with artistic exploration. Within the earlier four-year timeframe, the focus was more constricted, with just two works emphasizing the use of microbes ``FUTURE FLORA"~\cite{ars2022futureflora} and cyanobacteria ``I’M HUMANITY"~\cite{ars2017humanity} and one specifically concentrating on mushrooms ``PROJECT ALIAS"~\cite{ars2019alias}, demonstrating a more limited engagement with the vast potential of biological art at that time.

\subsection{Data Analysis on ISEA}

\begin{figure}
\includegraphics[width=\textwidth]{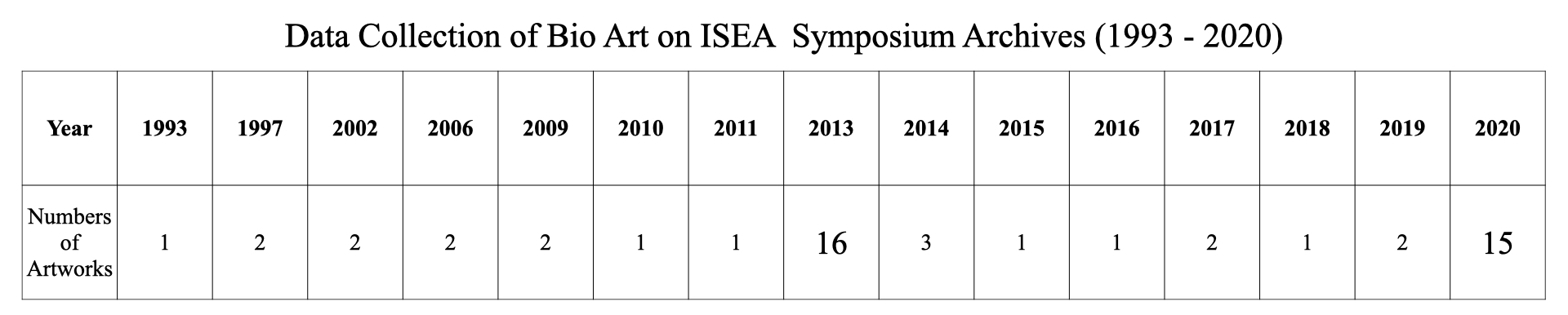}
\caption{The Bio Art Archives on ISEA Symposium from 1993 to 2020.
\copyright Table Credit by Authors.}
\Description{The Bio Art Archives on ISEA Symposium from 1993 to 2020.}
\label{isea}
\end{figure}

Our examination of the biological art archive on ISEA reveals a total of 52 works, cataloged by the years they were exhibited and published, as opposed to the year of their creation (see Figure~\ref{isea}). 2020 saw 15 biological art pieces added to the collection, while 2013 stands out with 16 entries, the highest count from 1993 to 2020.

The shift from 2013 to 2020 in the biological art archive from ISEA is marked by a significant evolution in thematic focus. By analyzing the motif of the listed sixteen works in 2013 and fifteen works in 2020, this study identifies a clear trajectory from anthropocentric to biocentric themes, reflecting a broader cultural and scientific engagement with environmental issues and the non-human world.

\begin{figure}
\includegraphics[width=\textwidth]{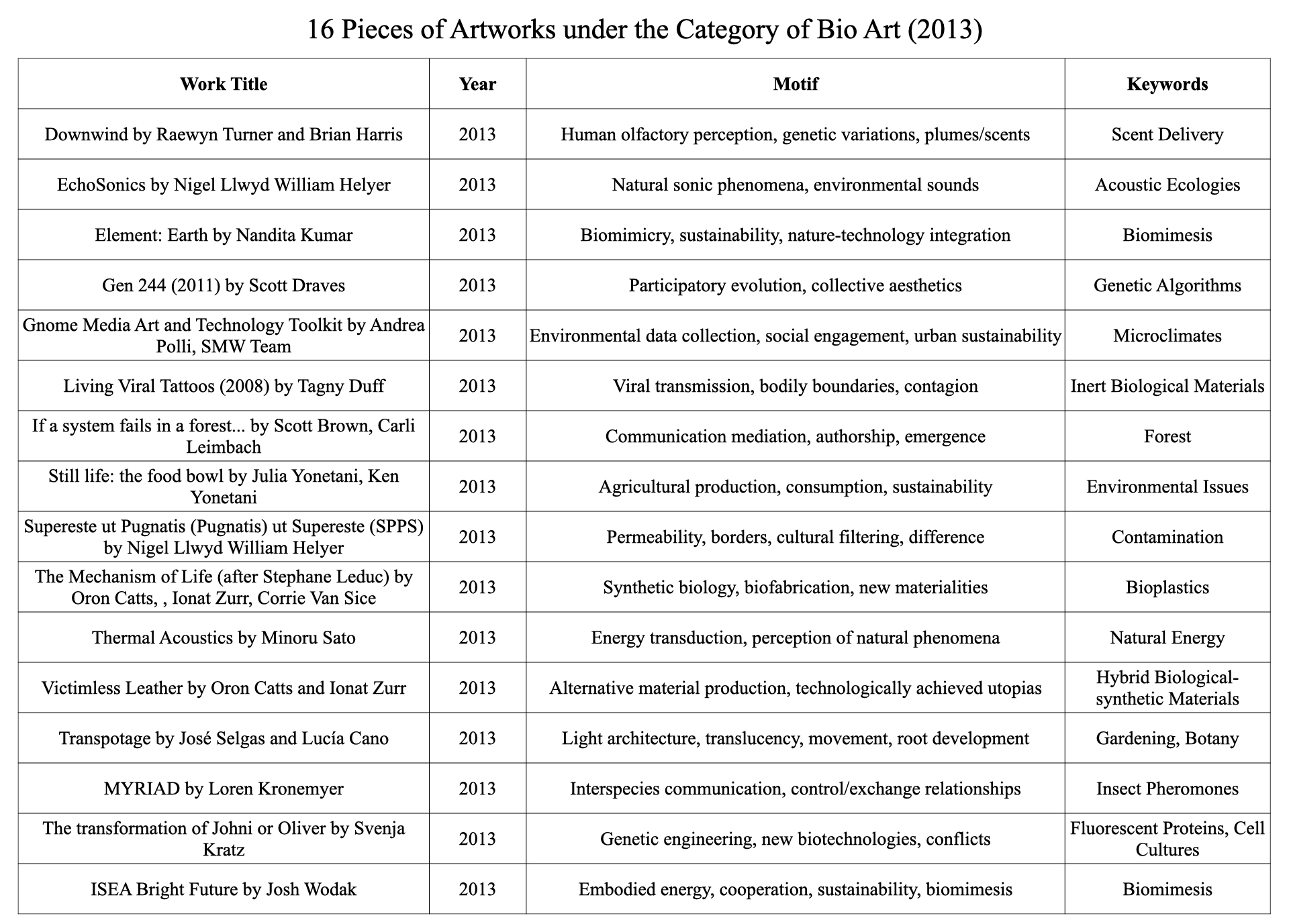}
\caption{The Biological Artworks Exhibited and Published in 2013 on ISEA.
\copyright Table Credit by Authors.}
\Description{The Biological Artworks Exhibited and Published in 2013 on ISEA.}
\label{isea13}
\end{figure}

The 2013 collection highlights artworks that emphasize humanity's deep-seated connection with the larger ecological matrix, challenging and moving beyond the entrenched distinctions between humans and the natural world from the keywords marked in our study (see Figure~\ref{isea13}). There is an acknowledgment of the integrated nature of humans within broader ecological systems, suggesting the beginnings of a shift toward a biocentric, or ecology-centered, perspective. These pieces incorporate botanical or microbial elements to varying extents and collectively work toward dismantling the notion of human exceptionalism, depicting humans as an integral component of, rather than separate from, the complex ecological systems and communities they artistically interpret.

\begin{figure}
\includegraphics[width=\textwidth]{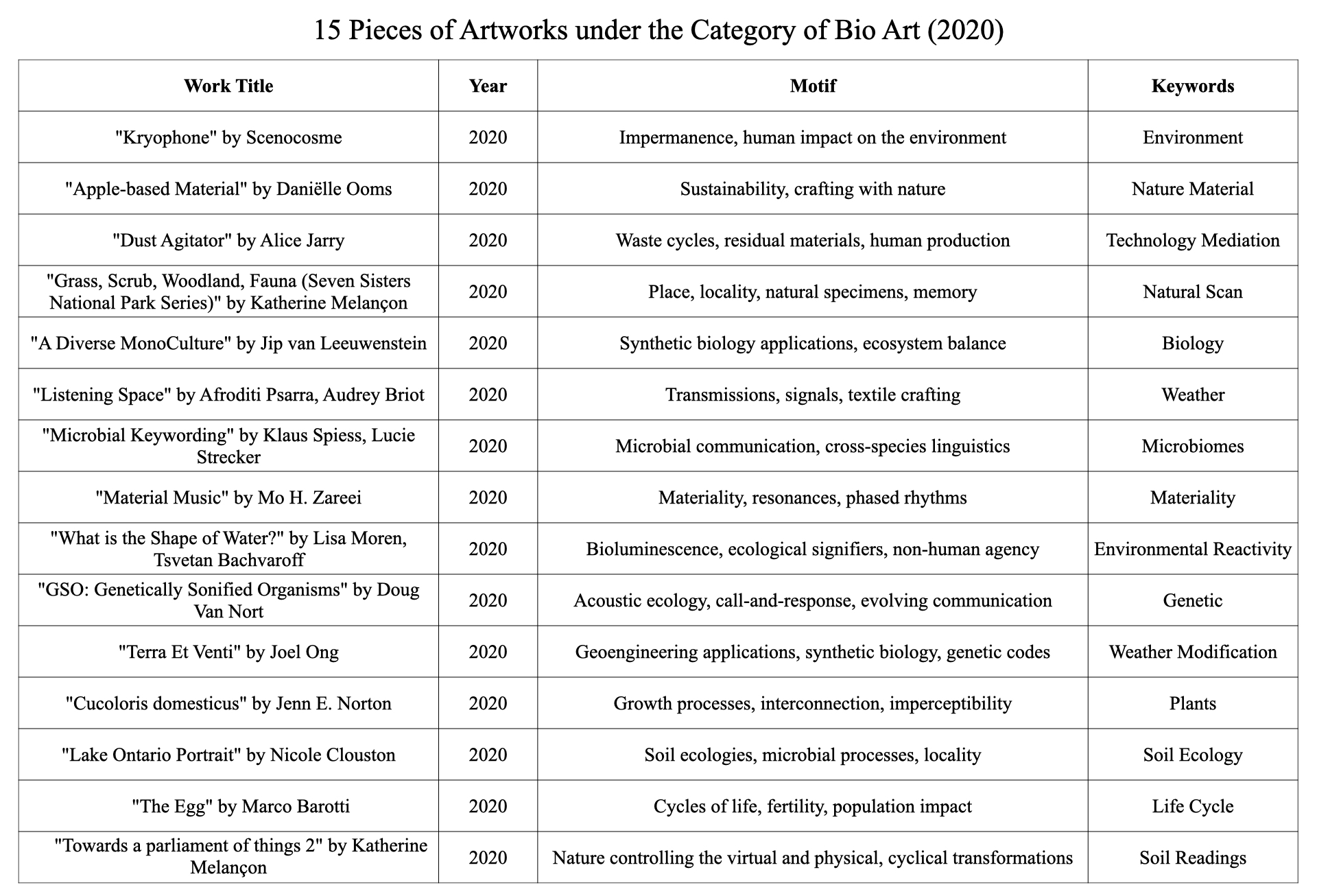}
\caption{The Biological Artworks Exhibited and Published in 2020 on ISEA.
\copyright Table Credit by Authors.}
\Description{The Biological Artworks Exhibited and Published in 2020 on ISEA.}
\label{isea20}
\end{figure}

In contrast, the 2020 collection presents works that explore both plant-focused and human-related themes (see Figure~\ref{isea20}). The interplay between human actions and ecological processes is pivotal to the concept and objective of these pieces. Employing elements such as the environment, natural materials, genetics, plants, soil ecology, and soil analysis, the primary aim of these artworks is to foster innovative forms of interaction, representation, and communication that go beyond the limitations of human perception and dominion in the growth of biology. These pieces are often rooted in human-related environmental concerns, like climate change, and offer a sophisticated exploration of the relationships between species set against the backdrop of urgent ecological challenges.

While the works from 2021 onward have not yet been archived, the transition in biological art from an anthropocentric to an ecocentric perspective, followed by an ecology-influenced human activity approach observed in the ISEA archives from 2013 to 2020, exemplifies the initial three stages of the BCTI. This shift accentuates the capacity of art to progress and mirrors the intricate interplay within ecological networks. 

Continuing from these observations from ARS ELECTRONICA and ISEA, the next anticipated phase in the evolution of biological art, where plants are expected to act as a catalyst for human interaction within the Metaverse—suggests a blending of biological and digital realms. This convergence is a testament to art's transformative power, transcending traditional media and entering a space where biological processes and technological infrastructures intersect. Despite being in digital spaces, plants could serve as dynamic interfaces or environmental sensors within virtual environments, enabling participants to experience the rhythms and cycles of nature in real-time. This could lead to cyber-ecology, where the organic and virtual boundaries are artistically interwoven to create a continuous feedback loop of influence and response between the two systems, affecting human gathering activities online or offline. As humans navigate these metatarsal spaces, their actions could be informed and even directed by live plants' growth patterns, health, and responses, making the ecological impact of human behavior more immediate and palpable.

\section{Discussion and Implications}

This research offers a comprehensive framework analyzing the evolution of biocentric ideologies in art. Our BCTI charts a course from anthropocentric to ecocentric perspectives, positioning art as an active participant within broader ecological systems. We provide a concrete methodology integrating biological components and ecological principles into artistic practice. Analyzing cultural, biological, semiotic, and aesthetic dimensions, we elucidate the expanding role of plants as dynamic subjects and mediums in art. Case studies demonstrate the migration of ecological art into digital realms like the Metaverse, redefining engagement. This research contributes to the discourse on art's role in fostering ecological consciousness and stewardship. It envisions a cultural shift emphasizing art’s influence within interconnected natural and virtual environments. Our interdisciplinary approach enriches comprehension of how artistic innovation can transcend boundaries, honoring the sanctity of all life forms.

\subsection{Sociocultural Ramifications of Emerging Perspectives in Art}

\textbf{Implications of plant-centered art}. Plant-centered art has been gaining prominence, reflecting a shift in how humans perceive and interact with the vegetal world. This type of art often integrates living plants or plant motifs to communicate messages about ecology, symbiosis, and the non-human agency of plants. The discussion around plant-centered art focuses on how it can change perceptions, foster empathy with the natural world, and highlight the importance of plants in our ecosystems. It challenges the anthropocentric view that has traditionally placed human beings at the center of the world and instead promotes a more biocentric or ecocentric perspective that recognizes the vital role of plants in sustaining life on Earth.

The plant-centered art, on a cultural level, can lead to greater environmental awareness and sensitivity toward the needs of ecosystems. It may encourage policies that prioritize the preservation of plant life and biodiversity. In an educational context, such art can serve as a tool to teach about the complex processes of photosynthesis, respiration, and the intricacies of plant communication and intelligence. Economically, the rise of plant-centred art might stimulate new industries and markets, including biophilic design, sustainable architecture, and eco-friendly art materials. Socially, it can foster a sense of community around environmental conservation efforts and translate into more grassroots activism.

\textbf{New paradigms and ideologies}. The emergence of new paradigms and ideologies, often influenced by advancements in science and technology, reshapes the landscape of art and culture. These new frameworks challenge the traditional binaries of nature/culture, organic/synthetic, and human/non-human. They propose a more integrated view of life where boundaries are permeable, and identities are complex and adaptive. The discussion involves how these ideologies can influence art and how societies organize, govern, and understand themselves in the natural world.

The implications of these new paradigms and ideologies are profound and far-reaching. Politically, they can lead to new forms of governance that are more inclusive of non-human entities and their rights. Ethically, they call for reconsidering our responsibilities to the environment and the non-human world, potentially leading to new legal and moral frameworks that recognize the intrinsic value of all forms of life. In terms of identity, these paradigms promote a more fluid and composite understanding of the self, which is interwoven with technology and the environment. This could result in more holistic approaches to education, mental health, and community building that acknowledge the interconnectedness of all beings. On a practical level, such ideologies can drive innovation in sustainable technologies and provoke shifts in consumer behavior towards more ecologically responsible products and practices.

\subsection{Theoretical Underpinnings of Biocentric Artistic Explorations}

The discussion of BCTI intertwines various scholarly perspectives to elucidate how artists challenge and redefine the interplay between technology, biology, and culture within human identity. Cultural anthropology, focusing on diversity and commonality across societies, sheds light on the profound ways artistic endeavors can mirror and mold our understanding of humanity's connection to the natural world. Stelarc's innovative ``Third Hand" epitomizes this dialogue by blurring the lines between the living and the synthetic, suggesting a future where human identity transcends biological roots, becoming a malleable construct enhanced by technological extensions.

In this exploration, the structuralist search for universal patterns in cultural symbols becomes a tool for artists to uncover the deep-seated connections that bind life forms within their cultural habitats. Such artistic interpretations could expose the foundational elements of biocentrism, highlighting the essential relationships that link various life forms and the human symbolic order. Meanwhile, Foucault's insights into biopolitics and the governance of life cast a critical eye on how biological processes are controlled and manipulated. Art that engages with topics like genetic engineering critiques the societal imposition on bodies and identity, echoing Foucault's concerns over institutionalized body regulation. Furthermore, Butler's theories on gender performativity inspire art that confronts entrenched gender norms and the often unquestioned link between biological sex and gender roles, supporting BCTI's emphasis on identity as a dynamic performance.

The burgeoning field of plant neurobiology is depicted in art that grants plants an active role within ecosystems, not merely as a backdrop but as engaged actors, thus aligning with BCTI's principles by uplifting the status of plant life and acknowledging their integral ecological functions. In biosemiotics, artists may render plants' complex and otherwise imperceptible communication methods into tangible experiences, deepening human appreciation for the rich interconnectedness of all life forms. Ecological aesthetics further underpins art that visualizes the elaborate exchanges between humans and the broader ecological community, prompting reflection on our environmental impact and interconnected existence.

Furthermore, concepts rooted in cybernetics, theories of control and chaos, liberty, and adaptationism forge new pathways for BCTI-influenced art, examining how biocentric creations morph in unpredictable ways, often eluding human control. Such art challenges human-centered perspectives and suggests a coevolutionary tapestry where humans, technology, and nature are bound in a reciprocal, ever-changing relationship. This dynamic system subverts anthropocentric norms and envisions a collaborative existence and mutual adaptation.

\subsection{Navigating the Tensions Between Platonism and Representation in Ecological Art}

According to Platonism, our sensory experiences of the world are imperfect and transient, which means that while we can perceive physical objects, we can never fully comprehend their true essence because they are just copies or imitations of their perfect Forms. While the Forms are the perfect, eternal, and unchanging blueprints of all things that exist. For example, we can see many individual plants (none of which are perfect), but the Form of ``Plant-ness" itself is an abstract, perfect concept that embodies the essence of what it means to be a plant. Even though we can't perceive this Form with our senses, we can understand and know it through reasoning and intellect.

This issue arises within the philosophical framework of Platonism, which recognizes the existence of abstract concepts or Forms that are understood to be perfect and immutable, existing beyond time and space in the dynamic biological world. It specifically affects artists of ecological art who strive to embody and communicate intricate ecological ideas—ideas that often involve abstract, systemic relationships and phenomena—within the physical medium of art, such as sustainability, geopolitics, biodiversity, climate change, habitat loss, and human-nature interaction. Thus, they encounter the challenge of reconciling Platonism between the concrete world of their artistic media and the abstract concepts they aim to convey.

To navigate this philosophical terrain, ecological artists must engage with both the visible and the invisible aspects of nature, such as to find a way to represent the physicality of the environment and the conceptual underpinnings that define its very existence. This involves translating the intangible aspects of ecology—the patterns, processes, and connections that are essential but not directly observable—into tangible artistic forms. Their work becomes a conduit for exploring the implications of human actions on the environment, inviting reflection on how these actions deviate from the natural orders and, in turn, how these spreading natural communities are affecting humans.

\subsection{Limitations and Future Works}

When considering new paradigms and ideologies, their integration into current societal norms and political frameworks is a significant limitation. These innovative ideas often face resistance from entrenched systems, for example, fossil fuel dependence, traditional agricultural practices, and consumerist economic models. Additionally, the discourse around these paradigms tends to circulate within academic and artistic communities, which limits their reach and impact on broader societal change.

Another notable limitation in plant-centered art is the difficulty in gauging its direct influence on the environmental consciousness of its audience, such as changes in recycling habits, reductions in carbon footprints, and shifts in consumer purchasing towards more sustainable products. Some decentralized research, for instance, planting and policy decisions, is also relevant to plants and communities\cite{xiang2023decentralized}. The subjective nature of the art means that interpretations and reactions can vary significantly among individuals, which poses a challenge in assessing the art's effectiveness in promoting ecological awareness or behavioral change. At the same time, collaborative projects that bring together artists, scientists, and educators could help create aesthetically compelling and scientifically grounded art while they have to follow or add novel thoughts to it. These collaborations could also facilitate the development of interactive and immersive art experiences that make ecological concepts more tangible and relatable to diverse audiences.

In addition, there is a limitation in how the artworks are collected and analyzed through these databases. The rationale behind the shift from human-centered to plant-centered aesthetics is now limited to the numerical fluctuation of biological artworks in different archives. However, this form of data collection relies upon ``Keyword," which might lead to the omission of other biological artworks that straddle between different categories or those related to biological art but not classified as such. Another possible loophole is that this transition from human-centered to plant-centered should not only be reflected through the increase in the number of biological artworks but also the number of these works that can be classified as plant-centered that have shown a clear impact of plant-centered aesthetics on human social behavior. Nevertheless, the lack of imagery and detailed descriptive information has made it difficult to make such classifications. Therefore, it requires a refined approach to data collection and classification in future research to improve our argument. Also, emerging technologies are prevalent, and VR archiving offers more opportunities and possibilities for media arts~\cite{gao2023vr}~\cite{guo2024echoes}.

\section{Conclusion}

By analyzing and detailing the biocentric creation transformation ideology, we've charted a course for artists to pivot from human-centered narratives to those that prioritize ecological and biotic considerations, advocating for the fusion of artistic practices with ecological principles. The diagrammatic framework provides artists with a tangible method for incorporating biological elements into their art, fostering an integration of ecological consciousness in their creative process. Our research expands the conversation around biocentric art, offering insights across social and biological dimensions, semiotics, ecological aesthetics, and their digital implications, particularly within the Metaverse. Meanwhile, as an educational tool, it guides new biological artists through conceptual and practical aspects of biocentric art, helping them to forge unique artistic identities.

We propose that biological art can transcend traditional mediums and flourish within digital environments like the Metaverse, effectively broadening the scope and influence of biocentric creations. This is part of a larger vision to foster a cultural shift towards ecologically integrated thinking and creating within the art community. By envisioning art as an active participant in shaping ecological and digital landscapes, we provide a foundation for artistic innovation that challenges traditional boundaries and embraces ecological narratives. Encouraging interdisciplinary exchange, we invite artists to enrich their work with insights from social sciences, biology, and technology, thus advancing the field of art and reinforcing its role in environmental advocacy and stewardship. We challenge artists to see themselves as integral participants in the ecological fabric, capable of influencing and being influenced by the world around them.

\bibliographystyle{ACM-Reference-Format}
\bibliography{main}

\end{document}